# Adaptive coupling of peridynamic and classical continuum mechanical models driven by broken bond/strength criteria for structural dynamic failure

JiuYi Li[1], ShanKun Liu[2], Fei Han[2,*], Yong Mei[1], YunHou Sun[1,*], FengJun Zhou[1]

(1. *National Defense Engineering Research Institute of Academy of Military Science of PLA, Beijing* 100850, *China*; 2. *State Key Laboratory of Structural Analysis for Industrial Equipment, Department of Engineering Mechanics, Dalian Universityl of Technology, Dalian* 116023, *China*)

Abstract: Peridynamics (PD) is widely used to simulate structural failure. However, PD models are time-consuming. To improve the computational efficiency, we developed an adaptive coupling model between PD and classical continuum mechanics (PD-CCM) based on the Morphing method [1], driven by the broken bond or strength criteria. We derived the dynamic equation of the coupled models from the Lagrangian equation and then the discretized finite element formulation. An adaptive coupling strategy was introduced by determining the key position using the broken bond or strength criteria. The PD subdomain was expanded by altering the value of the Morphing function around the key position. Additionally, the PD subdomain was meshed by discrete elements (DEs) (i.e., nodes were not shared between elements), allowing the crack to propagate freely along the boundary of the DE. The remaining subdomains were meshed by continuous elements (CEs). Following the PD subdomain expansion, the CEs were converted into DEs, and new nodes were inserted. The displacement vector and mass matrix were reconfigured to ensure calculation consistency throughout the solving process. Furthermore, the relationship between the expansion radius of the PD subdomain and the speed of crack propagation was also discussed. Finally, the effectiveness, efficiency, and accuracy of the proposed model were verified via three two-dimensional numerical examples.

# 1 Introduction

Issues related to structural failure have long been a prominent area of investigation within the field of mechanics. However, the classical continuum mechanics (CCM) model, which assumes that the structure being studied adheres to conditions of continuity and quadratic differentiability of deformation, encounters a challenge when attempting to solve the discontinuity problems of structure. This is due to the singularity problem encountered when employing the partial differential equation derived from the CCM model. To solve the singularity problem caused by discontinuous solutions, researchers have developed several crack-treatment methods, such as the extended finite element method [2,3], the element erosion method [4,5], and the phase field method [6,7]. Although these crack-treatment methods have been successfully used to analyze various fracture cases, obtaining continuous crack paths often requires using specialized crack-tracking techniques or mesh processing strategies. Consequently, the solution process becomes complicated when dealing with problems such as initiation, merging, and crack branching. Moreover, these methods [8,9], based on the theory of fracture mechanics, failed to provide a clear explanation of the mechanism behind crack formation. In 2000, Silling [10] proposed a new theory called peridynamics (PD), which differs from the CCM. The theory is based on the concept of non-local interaction. PD is anticipated to offer a new explanation for the mechanism of crack formation based on its current development.

PD employs integral equations to represent the equilibrium of forces. Unlike traditional partial differential equations, PD equations do not involve derivation operations. As such, PD describes both continuous and discontinuous deformations.

Naturally, PD can be used to simulate fracture behaviors such as initiation, propagation, and branching of cracks on structures [11-14]. However, the material points near the boundary, characterized by the PD model, lack a complete interaction neighborhood, resulting in a boundary effect [15]. Furthermore, PD assumes that the force on a material point is affected by all points in its neighborhood. Consequently, the numerical calculation cost of the PD model is relatively high [1]. To overcome these problems, a model coupling of PD-CCM can be employed to describe the computational domain. The domains (such as damage and fracture subdomains) in the structure are described by PD, whereas CCM describes the remaining parts of the structure. The scheme can embed the calculation subdomain of PD in the CCM calculation subdomain and impose constraints on the CCM subdomain boundary to avoid the boundary effect while also effectively reducing the computational cost. Therefore, the coupling strategy can exploit both theories while mitigating their disadvantages.

In recent years, researchers have proposed various strategies for PD-CCM coupling. Kilic and Madenci [16] realized PD-CCM coupling by introducing an overlapping subdomain of PD-CCM into the calculation subdomain of CCM and solving PD-CCM in this subdomain. Agwai et al. [17] proposed a sub-modeling approach for PD-CCM coupling. They used CCM to describe the undamaged subdomains of the structure and solved it using the finite element method (FEM) while using the PD model to describe the sub-modeling subdomain. Liu and Hong [18] proposed PD-CCM coupling using the interface element method. This method introduces the concept of a coupling force into the interface element to establish the mutual coupling of PD-CCM. Li [19] proposed a PD-FEM coupling method to address the dynamic problems of solid mechanics in crack propagation. Using the first-order Taylor expansion technique, this method derives an implicit PD formulation from the bond-based pairwise force described as a linear function of the displacements. The equivalent incremental equations of the PD and finite element methods are obtained based on the Newmark and Newton-Raphson schemes. Shen [20] proposed a new PD-FEM coupling model based on the implicit scheme. The displacements of a two-dimensional bar and a cantilever beam using the model are compared with theoretical solutions and FEM. The crack intersection in a slab is simulated and compared with other methods. Numerical predictions of the mode I fracture in a three-point bending beam agree well with the experimental results. Ongaro [21] presents a neglected problem in this type of PD-CCM coupling model: the lack of global balance. The problem was determined to be due to the presence of high-order derivatives of displacements in the coupling domain. Galvanetto et al. [22] proposed a coupling strategy that employs common nodes instead of overlapping subdomains. First, the computational domain is divided into two subdomains: one is described by the PD model and the other by the CCM model. The coupling between these two models is accomplished by making the PD bond forces act exclusively on the PD nodes and the stresses in CCM act only on the finite element nodes. In the above coupling strategies, establishing the coupling balance equation for the points of the interaction boundary, or overlapping domain, is necessary. This method often increases the computational complexity while leading to computational errors that cannot be ignored.

Lubineau and Azdoud et al. [1,23] proposed Morphing coupling. In this method, the PD to CCM subdomain transition relies only on the definition of material parameters in the models of both theories. A weighting function is employed to gradually change the value of material parameters in the model, controlling the results of the free transformation of the two models. The energy density equivalence of each point restricts the transformation without introducing any additional constraints to establish the transitioning relationship between subdomains. This method has been employed in PD-CCM coupling and simulation for static or quasi-static problems [23,24], and its effectiveness and reliability have been verified from theoretical and practical results. Han and Liu [25] applied the Morphing method to analyze structural dynamic failure in 2020. In their proposed dynamic hybrid local/non-local continuum model, the PD model must be arranged on a specified local subdomain before simulation. However, these fixed subdomains may restrict crack propagation or lead to increased computational costs. Furthermore, Azdoud et al. [26] proposed the adaptive expansion of PD subdomains driven by broken bonds to replace the strategy of fixed subdomains. The initial PD subdomain at the susceptible cracking subdomain is required to implement this method. However, when the structure is complicated, determining the location of crack initiation in advance is difficult, and the initial PD subdomain cannot be set in advance. Wang et al. [27] proposed a scheme in which the strength criterion drives the adaptive expansion of the PD subdomain to more flexibly expand it. The strength criterion of CCM replaces the broken bond criterion of PD in this scheme. Once the material strength is reached at certain positions, PD subdomains are activated near these positions to describe the fracture process.

The PD-CCM coupling model based on the Morphing method is a continuous research topic. Further, this paper focuses on the dynamic model of the adaptive coupling of the PD-CCM, the keyword being "adaptive." That is, the PD subdomain is not fully pre-assigned at the beginning of the calculation. As time passes, the adaptive coupling of the PD-CCM model can make the PD subdomain expand automatically with the crack. Thus, it can effectively predict the dynamic fracture results and minimize the use of the PD subdomain, reducing the calculation cost as much as possible. Therefore, achieving the adaptive expansion of PD subdomains driven by the broken bond and strength criteria in dynamic models is of potential research and application significance. This study aims to derive the dynamic formula of coupled PD-CCM using the Lagrangian equation and Morphing method. The adaptive coupling of the PD-CCM driven by the broken bond and strength criteria is applied to solve the dynamic failure problem of elastic-brittle materials and fill the research gap in this field. The remainder of this paper is organized as follows: Section 2 briefly reviews the constitutive model of bond-based PD, while Section 3 derives the finite element discrete equation of the coupled PD-CCM dynamic formula using the Lagrangian equation and the Morphing method. Section 4 introduces a numerical algorithm to achieve the adaptive expansion of the PD subdomain along crack paths by controlling the Morphing function. Section 5 tests the accuracy, effectiveness, and efficiency of the adaptive coupling of the PD-CCM to solve the dynamic failure problem of elastic-brittle materials using three benchmark examples. Finally, Section 6 summarizes the conclusions.

## 2 Bond-based PD constitutive model

Silling [10] proposed a bond-based PD model based on the non-local interaction in 2000 to avoid the singularity of the CCM model based on the continuity assumption when encountered the discontinuity issue. In PD, the domain $\Omega$, $\Omega \subset R^d\,(d=1,2,3)$ is considered as a continuum comprising a series of points $\boldsymbol{x}$. $H_\delta(\boldsymbol{x}) \subset \Omega$ is a spherical domain with radius $\delta$ centered at point $\boldsymbol{x}$, which is called the family of point $\boldsymbol{x}$ (Fig. 1).

$$\boldsymbol{f}(\boldsymbol{x},\boldsymbol{x}',t) = \hat{\boldsymbol{f}}(\boldsymbol{x},\boldsymbol{x}',t) - \hat{\boldsymbol{f}}(\boldsymbol{x}',\boldsymbol{x},t) \tag{1}$$

where $\hat{\boldsymbol{f}}(\boldsymbol{x}',\boldsymbol{x},t)$ denotes the partial interaction of point $\boldsymbol{x}'$ on point $\boldsymbol{x}$ at time $t$, and $\hat{\boldsymbol{f}}(\boldsymbol{x},\boldsymbol{x}',t)$ denotes the partial interaction of point $\boldsymbol{x}$ on point $\boldsymbol{x}'$ at time $t$ [11].

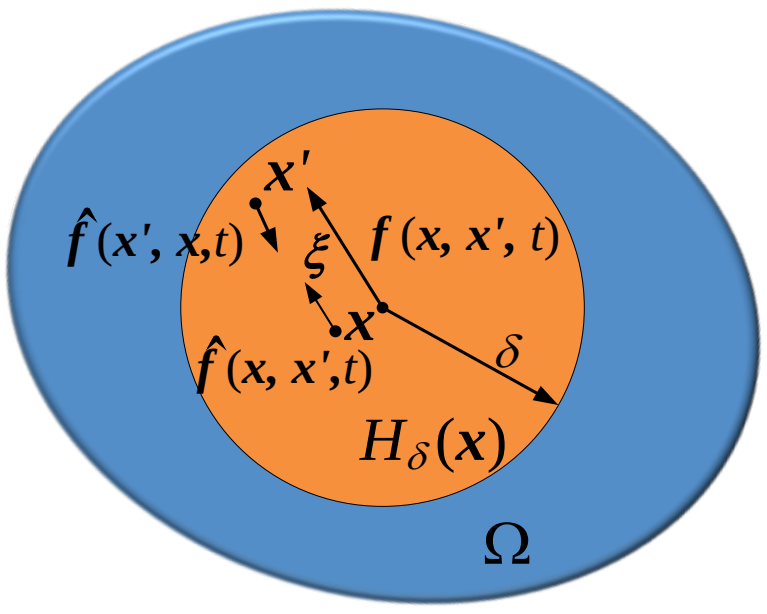

Fig. 1 Schematic of bond interaction in bond-based PD.

The constitutive model of bond-based PD under the assumptions of linear elasticity,

small deformation, and homogeneity can be expressed as follows:

$$\hat{\boldsymbol{f}}(\boldsymbol{x}', \boldsymbol{x}, t) = \frac{1}{2}\mathbf{C}(\boldsymbol{x}, \boldsymbol{\xi}) \cdot \boldsymbol{\eta}(\boldsymbol{x}', \boldsymbol{x}, t) \tag{2}$$

where $\boldsymbol{\eta}(\boldsymbol{x}', \boldsymbol{x}, t) = \boldsymbol{u}(\boldsymbol{x}', t) - \boldsymbol{u}(\boldsymbol{x}, t)$, $\boldsymbol{u}$ denotes displacement, and $\mathbf{C}(\boldsymbol{x}, \boldsymbol{\xi})$ is the micro-modulus tensor function defined as:

$$\mathbf{C}(\boldsymbol{x}, \boldsymbol{\xi}) = c(\boldsymbol{x}, \boldsymbol{\xi}) \boldsymbol{\xi} \otimes \boldsymbol{\xi} \tag{3}$$

where $c(\boldsymbol{x}, \boldsymbol{\xi})$ is the micro-modulus coefficient.

In PD, once the bond stretch $s$ is larger than a critical value $s_{crit}$, the bond breaks irreversibly [28]. The stretch $s$ of bond $\boldsymbol{\xi}$ is as follows:

$$s = \frac{\|\boldsymbol{\eta} + \boldsymbol{\xi}\| - \|\boldsymbol{\xi}\|}{\|\boldsymbol{\xi}\|} \tag{4}$$

The bond failure is achieved by introducing a history-dependent scalar-valued function $\mu(\boldsymbol{\xi}, t)$. For elastic-brittle materials, $\mu(\boldsymbol{\xi}, t)$ is defined as:

$$\mu(\boldsymbol{\xi}, t) = \begin{cases} 1 & \text{if} \quad s(\boldsymbol{\xi}, t') < s_{crit} \quad \text{for all} \quad 0 \le t' \le t \\ 0 & \qquad\qquad \text{otherwise} \end{cases} \tag{5}$$

where $s_{crit}$ is the critical stretch of the bond $\boldsymbol{\xi}$. When the stretch $s(\boldsymbol{\xi}, t')$ of the bond $\boldsymbol{\xi}$ is less than $s_{crit}$, $\mu = 1$ indicates that the bond $\boldsymbol{\xi}$ is intact. Otherwise, $\mu = 0$ indicates that the bond $\boldsymbol{\xi}$ is completely broken and cannot be recovered.

Then, the effective damage at point $\boldsymbol{x}$ can be defined as follows:

$$\phi(\boldsymbol{x}, t) = 1 - \frac{\int_{H_\delta(\boldsymbol{x})} \mu(\boldsymbol{\xi}, t) w_{crit} \mathrm{d}V_\xi}{\int_{H_\delta(\boldsymbol{x})} w_{crit} \mathrm{d}V_\xi} \tag{6}$$

where the critical fracture dissipation energy $w_{crit}$ of a single bond can be expressed as:

$$w_{crit} = \frac{1}{2} c(\boldsymbol{x}, \boldsymbol{\xi}) s_{crit}^2 \|\boldsymbol{\xi}\|^4 \tag{7}$$

# 3 Coupling model between PD-CCM

## 3. 1 Lagrangian description of the coupled model

As shown in Fig. 2, the domain $\Omega$ is divided into three subdomains: $\Omega_1$, $\Omega_2$, and $\Omega_m$. The relationship between the subdomains and domain can be described as follows: $\Omega = \Omega_1 \cup \Omega_2 \cup \Omega_m$, $\Omega_1 \cap \Omega_2 = \varnothing$, $\Omega_1 \cap \Omega_m = \varnothing$, and $\Omega_2 \cap \Omega_m = \varnothing$. $\Omega_2$ is completely embedded in $\Omega_m$, and $\Omega_m$ is completely embedded in $\Omega_1$, i.e., $\partial\Omega_2 \cap \partial\Omega_m = \varnothing$ and $\partial\Omega \subset \partial\Omega_1$. The displacement $\bar{\boldsymbol{u}}$ is imposed on part $\Gamma_{\boldsymbol{u}}$ of boundary $\partial\Omega$, the traction $\bar{\boldsymbol{F}}$ is imposed on part $\Gamma_{\boldsymbol{F}}$ of boundary $\partial\Omega$, and $\boldsymbol{n}$ is an outward unit normal vector. Additionally, the domain $\Omega$ is subjected to external body forces denoted by $\boldsymbol{b}$. The PD model is implemented in subdomain $\Omega_2$, enabling the spontaneous initiation and propagation of cracks. The continuum mechanical model is implemented in subdomain $\Omega_1$ so that traction can be imposed. The two models have a smooth transition on subdomain $\Omega_m$.

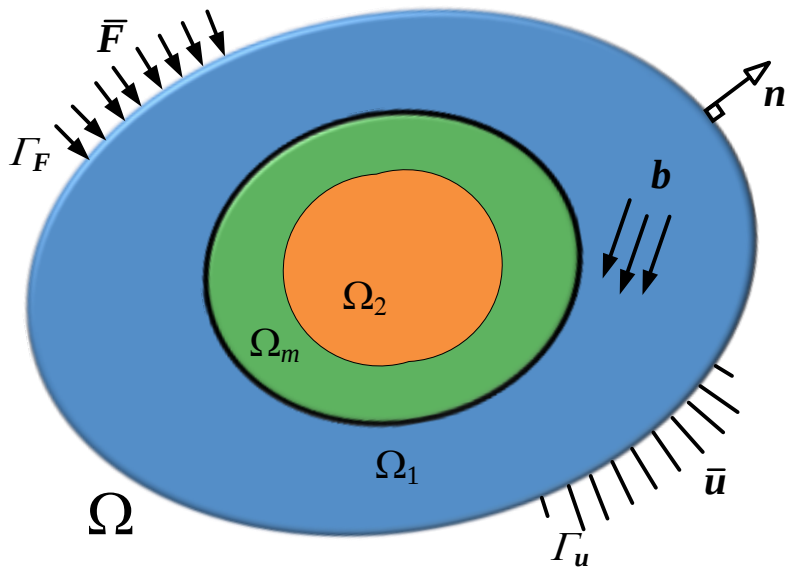

Fig. 2 Domain $\Omega$, comprising the continuum mechanical subdomain $\Omega_1$, the PD subdomain $\Omega_2$, and the transition subdomain $\Omega_m$.

The PD-CCM coupled system above can be represented by the Lagrangian equation in Eq. (8):

$$\frac{d}{dt}\left(\frac{\partial L(\boldsymbol{u},\dot{\boldsymbol{u}},t)}{\partial \dot{\boldsymbol{u}}}\right) - \frac{\partial L(\boldsymbol{u},\dot{\boldsymbol{u}},t)}{\partial \boldsymbol{u}} = \boldsymbol{0} \tag{8}$$

where $\boldsymbol{u}$ and $\dot{\boldsymbol{u}}$ denote the generalized displacement and generalized velocity vectors of the system, respectively. $L(\boldsymbol{u},\dot{\boldsymbol{u}},t)$ is the Lagrangian function, which can be defined as follows:

$$L(\boldsymbol{u},\dot{\boldsymbol{u}},t)=T-U \tag{9}$$

where $T$ denotes the total kinetic energy of the system.

$$T=\frac{1}{2}\int_{\Omega}\rho(\boldsymbol{x})\dot{\boldsymbol{u}}^2(\boldsymbol{x},t)dV_x \tag{10}$$

where $\rho(\boldsymbol{x})$ denotes the density at point $\boldsymbol{x}$. For the coupled PD-CCM system in Fig. 2, the total potential energy $U$ can be expressed as [25]:

$$U=\int_{\Omega}W_{\text{CCM}}(\boldsymbol{x},t)d\Omega+\int_{\Omega}W_{\text{PD}}(\boldsymbol{x},t)d\Omega-\int_{\Omega}W_{\boldsymbol{b}}(\boldsymbol{x},t)dV_{\boldsymbol{x}}-\int_{\Gamma_{\boldsymbol{F}}}W_{\bar{\boldsymbol{F}}}(\boldsymbol{x},t)dS_{\boldsymbol{x}} \tag{11}$$

where $W_{\boldsymbol{b}}(\boldsymbol{x},t)$ and $W_{\bar{\boldsymbol{F}}}(\boldsymbol{x},t)$ are the external works of body force $\boldsymbol{b}$ and traction $\bar{\boldsymbol{F}}$, respectively, defined as follows:

$$W_b(\boldsymbol{x},t)=\boldsymbol{u}(\boldsymbol{x},t)\boldsymbol{b}(\boldsymbol{x},t) \tag{12}$$

$$W_{\bar{F}}(\boldsymbol{x},t)=\boldsymbol{u}(\boldsymbol{x},t)\bar{\boldsymbol{F}}(\boldsymbol{x},t) \tag{13}$$

$W_{\text{CCM}}(\boldsymbol{x},t)$ and $W_{\text{PD}}(\boldsymbol{x},t)$ in Eq. (11) are the deformation energy densities at point $\boldsymbol{x}$ as described by continuum mechanics and PD, respectively. $W_{\text{CCM}}(\boldsymbol{x},t)$ and $W_{\text{PD}}(\boldsymbol{x},t)$ are defined separately by Eqs. (14) and (15), respectively.

$$W_{\text{CCM}}(\boldsymbol{x},t)=\frac{1}{2}\boldsymbol{\varepsilon}(\boldsymbol{x},t):\mathbf{E}(\boldsymbol{x}):\boldsymbol{\varepsilon}(\boldsymbol{x},t) \tag{14}$$

$$W_{\text{PD}}(\boldsymbol{x},t)=\frac{1}{4}\int_{H_\delta(\boldsymbol{x})}\boldsymbol{\eta}(\boldsymbol{x}',\boldsymbol{x},t):\frac{\mathbf{C}(\boldsymbol{x}',\xi)+\mathbf{C}(\boldsymbol{x},\xi)}{2}:\boldsymbol{\eta}(\boldsymbol{x}',\boldsymbol{x},t)dV_{\boldsymbol{x}'} \tag{15}$$

where $\boldsymbol{\varepsilon}(\boldsymbol{x},t)$ denotes the strain tensor of point $\boldsymbol{x}$ at time $t$, and $\mathbf{E}(\boldsymbol{x})$ is the stiffness tensor of CCM at point $\boldsymbol{x}$. The subdomains are assigned by introducing a Morphing function $\alpha(\boldsymbol{x})$ defined on the whole domain $\Omega$, and the range of the value $\alpha(\boldsymbol{x})$ is between [0,1]. Let $c(\boldsymbol{x},\xi)=\alpha(\boldsymbol{x})c^0(\|\xi\|)$ in Eq. (3), where $c^0(\|\xi\|)$ is a scalar-valued function. $\mathbf{C}(\boldsymbol{x},\xi)$ can be redefined as:

$$\mathbf{C}(\boldsymbol{x},\xi)=\alpha(\boldsymbol{x})c^0(\|\xi\|)\xi\otimes\xi \tag{16}$$

$\mathbf{E}(\boldsymbol{x})$ and $\mathbf{C}(\boldsymbol{x},\xi)$ are independent of time $t$. According to the constraint conditions of pointwise energy conservation [29], the definition of stiffness tensor $\mathbf{E}(\boldsymbol{x})$ at any point $\boldsymbol{x}$ in the coupled system can be defined as:

$$\mathbf{E}(\boldsymbol{x})=\mathbf{E}^{\mathbf{0}}-\int_{H_\delta(\boldsymbol{x})}\frac{\alpha(\boldsymbol{x}')+\alpha(\boldsymbol{x})}{2}c^0(\|\xi\|)\xi\otimes\xi\otimes\xi\otimes\xi dV_{\boldsymbol{x}'} \tag{17}$$

From Eq. (17), it can be deduced that $\mathbf{E}(\boldsymbol{x})=\mathbf{E}^{\mathbf{0}}$ when $\alpha(\boldsymbol{x}')=0, \forall \boldsymbol{x}' \in H_\delta(\boldsymbol{x})$, the coupled system of PD-CCM (i.e., Eq. (11)) degenerates into a pure continuum mechanical model; then once $\alpha(\boldsymbol{x}')=1, \forall \boldsymbol{x}' \in H_\delta(\boldsymbol{x})$, $\mathbf{E}(\boldsymbol{x})=\mathbf{0}$, the coupling system of PD-CCM degenerates into a pure PD model. In summary, the subdomains $\Omega_1$, $\Omega_2$, and $\Omega_m$ can be precisely defined based on the Morphing function $\alpha(\boldsymbol{x})$ as:

$$\begin{cases} \Omega_1 = \{ \boldsymbol{x} \mid \alpha(\boldsymbol{x}')=0,\ \forall \boldsymbol{x}' \in H_\delta(\boldsymbol{x}) \} \\ \Omega_2 = \{ \boldsymbol{x} \mid \alpha(\boldsymbol{x}')=1,\ \forall \boldsymbol{x}' \in H_\delta(\boldsymbol{x}) \} \\ \Omega_m = \{ \boldsymbol{x} \mid \exists \boldsymbol{x}' \in H_\delta(\boldsymbol{x}) \text{ such that } 0<\alpha(\boldsymbol{x}')<1 \} \end{cases} \tag{18}$$

## 3. 2 Discretization of the Lagrangian equations

We discretized the entire domain $\Omega$ into a finite number of elements $V_i (i=1,2,\ldots,n)$, where $n$ is the number of elements. These elements do not overlap and share nodes with adjacent elements, namely, $\Omega = V_1 \cup V_2 \cup \cdots \cup V_n$. Without a loss of generality, we define $\Omega_1 = V_1 \cup V_2 \cup \cdots \cup V_{n'}$, $\Omega_m = V_{n'+1} \cup V_{n'+2} \cup \cdots \cup V_{n''}$, and $\Omega_2 = V_{n''+1} \cup V_{n''+2} \cup \cdots \cup V_n$, where $1<n'<n''<n$. Generalized displacement vector $\boldsymbol{u}$ and generalized velocity vector $\dot{\boldsymbol{u}}$ can be expressed by the displacement vector $\tilde{\boldsymbol{u}}(t)$ and velocity vector $\dot{\tilde{\boldsymbol{u}}}(t)$ of the nodes as:

$$\boldsymbol{u} = \mathbf{N}(\boldsymbol{x})\tilde{\boldsymbol{u}}(t), \quad \dot{\boldsymbol{u}} = \mathbf{N}(\boldsymbol{x})\dot{\tilde{\boldsymbol{u}}}(t) \tag{19}$$

where $\mathbf{N}(\boldsymbol{x})$ denotes a piecewise continuous generalized shape function defined over the entire domain $\Omega$:

$$\mathbf{N}(\boldsymbol{x})\big|_{\boldsymbol{x}\in V_i} = \mathbf{N}_i(\boldsymbol{x})\mathbf{R}_i,\ \forall V_i \in \Omega,\ i=1,2,\ldots,n \tag{20}$$

where $\mathbf{R}_i$ is the transformation matrix through which the shape function matrix $\mathbf{N}_i(\boldsymbol{x})$ on $V_i$ is mapped to the corresponding position of the generalized shape function matrix $\mathbf{N}(\boldsymbol{x})$.

Substituting Eq. (8) into Eq. (9), both ends of the equations are multiplied by a time-independent and arbitrarily small function $\delta\boldsymbol{u}(\boldsymbol{x}) = \mathbf{N}(\boldsymbol{x})\boldsymbol{\kappa}$, where $\boldsymbol{\kappa}$ is an arbitrarily small vector. Then, Eq. (8) can be rewritten as follows:

$$\frac{d}{dt}\left(\frac{\partial T}{\partial \dot{\boldsymbol{u}}}(\mathbf{N}(\boldsymbol{x})\boldsymbol{\kappa})\right)+\frac{\partial U}{\partial \boldsymbol{u}}(\mathbf{N}(\boldsymbol{x})\boldsymbol{\kappa})=\mathbf{0} \tag{21}$$

since $\boldsymbol{\kappa}$ is an arbitrarily small vector, the requirements for Eq. (21) to be established are:

$$\frac{d}{dt}\left(\frac{\partial T}{\partial \dot{\boldsymbol{u}}}\mathbf{N}(\boldsymbol{x})\right)+\frac{\partial U}{\partial \boldsymbol{u}}\mathbf{N}(\boldsymbol{x})=\mathbf{0} \tag{22}$$

with Eq. (23) and Eq. (24) given as:

$$\frac{\partial \dot{\boldsymbol{u}}}{\partial \dot{\tilde{\boldsymbol{u}}}}=\frac{\partial(\mathbf{N}(\boldsymbol{x})\dot{\tilde{\boldsymbol{u}}}(t))}{\partial \dot{\tilde{\boldsymbol{u}}}}=\mathbf{N}(\boldsymbol{x}) \tag{23}$$

$$\frac{\partial \boldsymbol{u}}{\partial \tilde{\boldsymbol{u}}}=\frac{\partial(\mathbf{N}(\boldsymbol{x})\tilde{\boldsymbol{u}}(t))}{\partial \tilde{\boldsymbol{u}}}=\mathbf{N}(\boldsymbol{x}) \tag{24}$$

Substituting Eq. (23) and Eq. (24) into Eq. (22), we obtain:

$$\frac{d}{dt}\left(\frac{\partial T}{\partial \dot{\tilde{\boldsymbol{u}}}}\right)+\frac{\partial U}{\partial \tilde{\boldsymbol{u}}}=\mathbf{0} \tag{25}$$

Let $\boldsymbol{u}_i(\boldsymbol{x},t)$ denote the displacement solution over element $V_i$ at time $t$, and $\dot{\boldsymbol{u}}_i(\boldsymbol{x},t)$ denote the velocity solution over element $V_i$ at time $t$, which are expressed as:

$$\boldsymbol{u}_i(\boldsymbol{x},t)=\mathbf{N}_i(\boldsymbol{x})\tilde{\boldsymbol{u}}_i(t),\ \dot{\boldsymbol{u}}_i(\boldsymbol{x},t)=\mathbf{N}_i(\boldsymbol{x})\dot{\tilde{\boldsymbol{u}}}_i(t) \tag{26}$$

where $\tilde{\boldsymbol{u}}_i(t)$ is the nodal displacement of element $V_i$ at time $t$ and $\dot{\tilde{\boldsymbol{u}}}_i(t)$ is the nodal velocity of element $V_i$ at time $t$. Meanwhile, we define a minimum element domain $(V_{\boldsymbol{x}}^1 \cup V_{\boldsymbol{x}}^2 \cup \ldots \cup V_{\boldsymbol{x}}^{h(\boldsymbol{x})})$ that encompasses the neighborhood $(H_\delta(\boldsymbol{x}) \cap \Omega)$ of point $\boldsymbol{x}$, where $\{V_{\boldsymbol{x}}^1, V_{\boldsymbol{x}}^2, \ldots, V_{\boldsymbol{x}}^{h(\boldsymbol{x})}\} \subset \{V_1, V_2, \ldots, V_n\}$ and $h(\boldsymbol{x})$ denote the number of elements in the minimum element domain. Furthermore, we convert the kinetic and potential energies integrated over the whole domain $\Omega$ into the sum of the integration results over each element. Thus, the finite element discrete format of the kinetic and potential energies of the coupled PD-CCM system can be expressed as follows:

$$T=\dot{\tilde{\boldsymbol{u}}}(t)^{\mathrm{T}}\frac{1}{2}\sum_{i=1}^{n}\left(\int_{V_i}\rho(\boldsymbol{x})\left[\mathbf{N}_i(\boldsymbol{x})\mathbf{R}_i\right]^{\mathrm{T}}\left[\mathbf{N}_i(\boldsymbol{x})\mathbf{R}_i\right]dV_{\boldsymbol{x}}\right)\dot{\tilde{\boldsymbol{u}}}(t) \tag{27}$$

$$
\begin{aligned}
U = \tilde{\boldsymbol{u}}^{\mathrm{T}}(t)\Bigg\{ & \frac{1}{2}\sum_{i=1}^{n}\int_{V_i}\left[\mathbf{HN}_i(\boldsymbol{x})\mathbf{R}_i\right]^{\mathrm{T}}\left[\mathbf{E}(\boldsymbol{x})\right]\left[\mathbf{HN}_i(\boldsymbol{x})\mathbf{R}_i\right]dV_{\boldsymbol{x}} + \\
& \frac{1}{4}\sum_{i=1}^{n}\sum_{j=1}^{h(\boldsymbol{x})}\int_{V_i}\int_{V_{\boldsymbol{x}}^j} c^0\left(\|\boldsymbol{\xi}\|\right)\frac{\alpha(\boldsymbol{x}')+\alpha(\boldsymbol{x})}{2}\left[\mathbf{N}_j(\boldsymbol{x}')\mathbf{R}_j - \mathbf{N}_i(\boldsymbol{x})\mathbf{R}_i\right]^{\mathrm{T}} \\
& \left[\boldsymbol{\xi}\otimes\boldsymbol{\xi}\right]\left[\mathbf{N}_j(\boldsymbol{x}')\mathbf{R}_j - \mathbf{N}_i(\boldsymbol{x})\mathbf{R}_i\right]dV_{\boldsymbol{x}'}dV_{\boldsymbol{x}}\Bigg\}\tilde{\boldsymbol{u}}(t) + \\
& \tilde{\boldsymbol{u}}^{\mathrm{T}}(t)\left\{\sum_{i=1}^{n}\int_{V_i}\left[\mathbf{N}_i(\boldsymbol{x})\mathbf{R}_i\right]^{\mathrm{T}}\left\{\boldsymbol{b}(\boldsymbol{x},t)\right\}dV_{\boldsymbol{x}} + \sum_{i=1}^{n}\int_{\Gamma_i}\left[\mathbf{N}_i(\boldsymbol{x})\mathbf{R}_i\right]^{\mathrm{T}}\left\{\bar{\boldsymbol{F}}(\boldsymbol{x},t)\right\}dS_{\boldsymbol{x}}\right\}
\end{aligned} \tag{28}
$$

where $\mathbf{H}$ denotes the differential operator matrix [25], and $V_{\boldsymbol{x}}^j$ , $(j = 1,2,\ldots,h(\boldsymbol{x}))$ denotes the interaction element in the neighborhood of point $\boldsymbol{x}$. By substituting Eq. (27) and Eq. (28) into Eq. (25), we obtain Eq. (29):

$$
\mathbf{M}\ddot{\tilde{\boldsymbol{u}}}(t) + \mathbf{K}\tilde{\boldsymbol{u}}(t) = \boldsymbol{F}(t) \tag{29}
$$

where $\ddot{\tilde{\boldsymbol{u}}}(t)$ denotes the nodal acceleration and $\mathbf{M}$ , $\mathbf{K}$ , and $\boldsymbol{F}(t)$ denote the mass matrix, stiffness matrix, and load vector, respectively, which can be expressed as follows:

$$
\mathbf{M} = \sum_{i=1}^{n}\int_{V_i}\rho(\boldsymbol{x})\left[\mathbf{N}_i(\boldsymbol{x})\mathbf{R}_i\right]^{\mathrm{T}}\left[\mathbf{N}_i(\boldsymbol{x})\mathbf{R}_i\right]dV_{\boldsymbol{x}} \tag{30}
$$

$$
\begin{aligned}
\mathbf{K} = & \sum_{i=1}^{n}\int_{V_i}\left[\mathbf{HN}_i(\boldsymbol{x})\mathbf{R}_i\right]^{\mathrm{T}}\left[\mathbf{E}(\boldsymbol{x})\right]\left[\mathbf{HN}_i(\boldsymbol{x})\mathbf{R}_i\right]dV_{\boldsymbol{x}} + \\
& \frac{1}{2}\sum_{i=1}^{n}\sum_{j=1}^{h(\boldsymbol{x})}\int_{V_i}\int_{V_{\boldsymbol{x}}^j} c^0\left(\|\boldsymbol{\xi}\|\right)\frac{\alpha(\boldsymbol{x}')+\alpha(\boldsymbol{x})}{2}\left[\mathbf{N}_j(\boldsymbol{x}')\mathbf{R}_j - \mathbf{N}_i(\boldsymbol{x})\mathbf{R}_i\right]^{\mathrm{T}} \\
& \left[\boldsymbol{\xi}\otimes\boldsymbol{\xi}\right]\left[\mathbf{N}_j(\boldsymbol{x}')\mathbf{R}_j - \mathbf{N}_i(\boldsymbol{x})\mathbf{R}_i\right]dV_{\boldsymbol{x}'}dV_{\boldsymbol{x}}
\end{aligned} \tag{31}
$$

$$
\boldsymbol{F}(t) = \sum_{i=1}^{n}\int_{V_i}\left[\mathbf{N}_i(\boldsymbol{x})\mathbf{R}_i\right]^{\mathrm{T}}\left\{\boldsymbol{b}(\boldsymbol{x},t)\right\}dV_{\boldsymbol{x}} + \sum_{i=1}^{n}\int_{\Gamma_i}\left[\mathbf{N}_i(\boldsymbol{x})\mathbf{R}_i\right]^{\mathrm{T}}\left\{\bar{\boldsymbol{F}}(\boldsymbol{x},t)\right\}dS_{\boldsymbol{x}} \tag{32}
$$

where $[\cdot]$ and $\{\cdot\}$ denote a matrix and a vector, respectively, and $\Gamma_i$ is the boundary of $V_i$ .

At the discretization level, the entire domain $\Omega$ is meshed by finite elements, where the continuum mechanical subdomain $\Omega_1$ and transition subdomain $\Omega_m$ are meshed by CEs, and the PD subdomain $\Omega_2$ is meshed by DEs. Unlike CEs, the nodes and edges are not shared between adjacent DEs. This spatial discretization method was successfully applied by Han and Li [26,30] in PD simulations.

# 4 Numerical implementation

## 4. 1 Adaptive algorithm for expanding PD subdomains

As the applied load on the structure increases, it undergoes an evolutionary process from elastic deformation to crack initiation and propagation. During this process, the elements in the structure are gradually damaged. We mark the centroid of each damaged element in each time step as a flag point and compose all marked centroid points into a flag-point set. By assigning the value of the Morphing function $\alpha$ (with a value less than or equal to 1, i.e., $\alpha \leq 1$) to the subdomain controlled by each flag point, a new PD subdomain is introduced. Consequently, the PD subdomain can be adaptively expanded. The manner of determining the flag-point set for adaptively introducing PD subdomains is divided into those driven by the "broken bond criterion" and those driven by the "strength criterion.”

### 4. 1. 1 Obtaining a flag-point set

#### 4. 1. 1. 1 Broken bond criterion-driven process

If the location of crack initiation can be predicted in advance, the broken bond criterion is selected to determine a flag-point set, and then the PD subdomain appears and expands spontaneously through the flag points guider. The broken bond criterion-driven process for determining flag points is that when a bond in the PD subdomain breaks, the two elements (here denoted as $V_p$ and $V_{p'}$) where the two ends of the bond are located are considered damaged. The centroids of the elements $V_p$ and $V_{p'}$ are marked as the flag points $\boldsymbol{p}$ and $\boldsymbol{p}'$ (see Fig. 3). The centroid points of all damaged elements in the current time step are computed as a flag-point set $\mathbb{C}$ defined as follows:

$$\mathbb{C} = \left\{ \boldsymbol{p}, \boldsymbol{p}' \mid \forall \boldsymbol{p}, \boldsymbol{p}' \in \Omega;\ \exists \boldsymbol{\xi}_{V_p,V_{p'}} \in \Omega \text{ and } \exists t', 0 < t' < t, \text{ such that } s(\boldsymbol{\xi}_{V_p,V_{p'}}, t') \geq s_{crit} \right\} \tag{33}$$

where $t'$ and $t$ denote the time, $\boldsymbol{\xi}_{V_p,V_{p'}}$ denotes the bond connected by taking a quadrature point from each of the damaged elements $V_{\boldsymbol{p}}$ and $V_{\boldsymbol{p}'}$, and $s(\boldsymbol{\xi}_{V_{\boldsymbol{p}},V_{\boldsymbol{p}'}},t')$ denotes the stretch of the bond $\boldsymbol{\xi}_{V_{\boldsymbol{p}},V_{\boldsymbol{p}'}}$ at time $t'$.

Remark: At the beginning of the simulation, some PD bonds must exist in the model to ensure that bond breakage can occur and guide the expansion of new PD subdomains. Therefore, the manner driven by the broken bond criterion for adaptive expansion of a PD subdomain must set some initial flag points $\boldsymbol{p}_0$ and $\boldsymbol{p}_0 \in \mathcal{C}$ at the location of crack initiation before the simulation.

#### 4. 1. 1. 2 The strength-criterion-driven process

Predicting the location of crack initiation is difficult owing to structural complexities. However, the strength criterion can be selected to spontaneously drive the appearance of new PD subdomains. Unlike the broken bond criterion-driven process, this process does not require setting the initial PD subdomain before the simulation. That is, at the beginning of the simulation, the CCM model describes the entire computational domain. However, the stress state at each element centroid point on the structure must be evaluated at each time step as the applied load increases. The element is considered damaged once the stress state of the centroid point reaches the critical stress state. Depending on the research, various strength criteria can be selected to evaluate whether the element is damaged. The criterion selected in this study is the generalized von Mises strength criterion, and its formula is:

$$\sigma_v = \sqrt{\frac{(\sigma_{11}-\sigma_{22})^2+(\sigma_{22}-\sigma_{33})^2+(\sigma_{33}-\sigma_{11})^2+6(\sigma_{12}+\sigma_{23}+\sigma_{31})^2}{2}} < \sigma_{crit} \qquad (34)$$

where $\sigma_{ij}\,(i,j=1,2,3)$ is a component of the stress tensor $\boldsymbol{\sigma}$, $\sigma_v$ denotes the calculated stress value at a point, and $\sigma_{crit}$ denotes the strength of the material. According to the generalized von Mises strength criterion, the centroid point of an element at which the evaluated stress value reaches the strength of von Mises material is denoted as $\boldsymbol{p}$. The corresponding element is called the damaged element and denoted as $V_{\boldsymbol{p}}$. Further, the centroid points of all damaged elements in the current time step comprise a flag-point set $\mathbb{C}$ , which is defined as follows:

$$\mathbb{C} = \{ \boldsymbol{p} \mid \forall \boldsymbol{p} \in \Omega;\ \exists t', 0 < t' < t,\ \text{such that}\ \sigma_v(\boldsymbol{p}, t') \geq \sigma_{crit} \} \tag{35}$$

where $\sigma_v(\boldsymbol{p}, t')$ denotes the value of von Mises stress calculated at the centroid point $\boldsymbol{p}$ of the element $V_{\boldsymbol{p}}$ at time $t'$.

### 4.1.2 The Morphing function $\alpha$

From the above, when the stress state of flag points $\boldsymbol{p}$ in $\mathbb{C}$ have reached the critical stress state $\sigma_{crit}$ or the deformation of the bonds related to $\boldsymbol{p}$ have reached the critical stretch $s_{crit}$, the continuum mechanical model must be converted into the PD model to simulate cracking because cracks may appear near point $\boldsymbol{p}$. Eq. (18) states that this conversion process can be achieved by assigning the Morphing function $\alpha$. The following procedure is used to determine the Morphing function $\alpha$ based on the flag-point set $\mathbb{C}$ without losing generality:

(1) At the beginning of the simulation, if the adaptive expansion process of the PD subdomain is driven by the broken bond criterion, then the initial flag point $\boldsymbol{p}_0$ exists in $\mathbb{C}$, and we can define the Morphing function $\alpha_{\boldsymbol{p}_0}(\boldsymbol{x})$ related to $\boldsymbol{p}_0$ as:

$$\alpha_{\boldsymbol{p}_0}(\boldsymbol{x}) = \begin{cases} 0 & \|\boldsymbol{x} - \boldsymbol{p}_0\| \geq r_2 \\ f_{\boldsymbol{p}}(\boldsymbol{x}) & \|\boldsymbol{x} - \boldsymbol{p}_0\| \in (r_1, r_2) \\ 1 & \|\boldsymbol{x} - \boldsymbol{p}_0\| \leq r_1 \end{cases} \quad \text{and} \quad f_{\boldsymbol{p}_0}(\boldsymbol{x}) \in (0,1),\ \forall \boldsymbol{x} \in \Omega \tag{36}$$

where $f_{\boldsymbol{p}_0}(\boldsymbol{x})$ is a scalar-valued smoothing conversion function related to the distance between $\boldsymbol{x}$ and $\boldsymbol{p}_0$. A cubic power function has been chosen in this study as follows:

$$f_{\boldsymbol{p}_0}(\boldsymbol{x}) = 1 + \frac{\left(\|\boldsymbol{x} - \boldsymbol{p}_0\| - r_1\right)^2 \cdot \left(2\|\boldsymbol{x} - \boldsymbol{p}_0\| - 3r_2 + r_1\right)}{\left(r_2 - r_1\right)^3} \tag{37}$$

where $r_2$ denotes the outer radius of the initial transition subdomain, and $r_1$ denotes the radius of the initial PD subdomain, which must cover the crack initiation domain. According to literature [1], to ensure the quality of the coupled model, it is necessary to satisfy $r_1 \geq \delta$ and $r_2 - r_1 \geq 2\delta$.

In contrast, if the adaptive expansion of the PD subdomain is driven by the strength criterion, then $\mathbb{C}$ is an empty point set at the beginning of simulation and the initial condition of the Morphing function $\alpha$ is given as follows:

$$\alpha(\boldsymbol{x}) = 0, \quad \forall \boldsymbol{x} \in \Omega \tag{38}$$

(2) During the simulation, the Morphing function $\alpha_{\boldsymbol{p}}(\boldsymbol{x})$ can be defined as:

$$\alpha_{\boldsymbol{p}}(\boldsymbol{x}) = \begin{cases} 0 & \|\boldsymbol{x}-\boldsymbol{p}\| \geq R_{\boldsymbol{p}} \\ f_{\boldsymbol{p}}(\boldsymbol{x}) & \|\boldsymbol{x}-\boldsymbol{p}\| \in (r_{\boldsymbol{p}}, R_{\boldsymbol{p}}) \\ 1 & \|\boldsymbol{x}-\boldsymbol{p}\| \leq r_{\boldsymbol{p}} \end{cases} \quad \text{and} \quad f_{\boldsymbol{p}}(\boldsymbol{x}) \in (0,1),\ \forall \boldsymbol{x} \in \Omega \tag{39}$$

Here, $f_{\boldsymbol{p}}(\boldsymbol{x})$ also selects the cubic power function shown in Eq. (37), where $r_{\boldsymbol{p}}$ and $R_{\boldsymbol{p}}$ are scalars and $R_{\boldsymbol{p}} > r_{\boldsymbol{p}}$. For any flag points $\boldsymbol{p}$ and $\boldsymbol{p}'$ in $\mathbb{C}$, $r_{\boldsymbol{p}}$ and $R_{\boldsymbol{p}}$ can be different from $r_{\boldsymbol{p}'}$ and $R_{\boldsymbol{p}'}$ when $\boldsymbol{p} \neq \boldsymbol{p}'$. In this study, $r_{\boldsymbol{p}}$ and $R_{\boldsymbol{p}}$ are selected as fixed values, where the value of $r_{\boldsymbol{p}}$ is the expansion radius of the PD subdomain used to describe the size of the adaptive expansion range of a PD subdomain. Notably, $r_{\boldsymbol{p}}$ must be greater than $\delta$ to ensure the appearance of a pure PD subdomain, and $R_{\boldsymbol{p}} - r_{\boldsymbol{p}}$ must be greater than $2\delta$ to ensure that the coupled model works well [1]. For all the numerical examples in this study, $r_{\boldsymbol{p}} = 2\delta$ and $R_{\boldsymbol{p}} = 4\delta$.

（3）If there are $n$ flag points $\boldsymbol{p}_i, i = 1, 2 \ldots, n$ in $\mathbb{C}$, then $n$ Morphing functions $\alpha_{\boldsymbol{p}_i}(\boldsymbol{x})$ are defined over the entire computational domain $\Omega$. These $\alpha_{\boldsymbol{p}_i}(\boldsymbol{x})$ overlap on $\Omega$. By taking the maximum value of these $\alpha_{\boldsymbol{p}_i}(\boldsymbol{x})$ at each point $\boldsymbol{x}$, we can conclusively define the uniform Morphing function $\alpha(\boldsymbol{x})$ on $\Omega$ as:

$$\alpha(\boldsymbol{x}) = \max\left\{\alpha_{\boldsymbol{p}_i}(\boldsymbol{x}) \mid \forall \boldsymbol{p}_i \in \mathbb{C}, i = 1, 2 \ldots, n\right\}, \quad \forall \boldsymbol{x} \in \Omega \tag{40}$$

Remark: The value of the Morphing function $\alpha(\boldsymbol{x})$ can be updated according to the change of $\mathbb{C}$ automatically, to convert the CCM model to the PD model; The value of $\alpha(\boldsymbol{x})$ is irreversible; if $\alpha(\boldsymbol{x}, t') = \alpha_1$, then $\alpha(\boldsymbol{x}, t) \geq \alpha_1$, $\forall t > t'$, where $t$ and $t'$ denote time.

### 4.1.3 Flowchart of the adaptive algorithm

This study proposes an adaptive coupling model of PD-CCM driven by the broken bond criterion or the strength criterion for structural dynamic failure. The model leverages the governing equation of hybrid PD-CCM described in Eq. (29), the constraint equation is given by Eq. (17), and the initial conditions are given by Eqs. (36) and (38). The Morphing function $\alpha$ is defined by Eqs. (39) and (40) according to a flag-point set $\mathbb{C}$, and the bond breaks are determined and recorded by Eq. (5). Together, the above formulation comprises a closed solution system that can be used for structural dynamic failure simulation. Specific implementation steps are as follows:

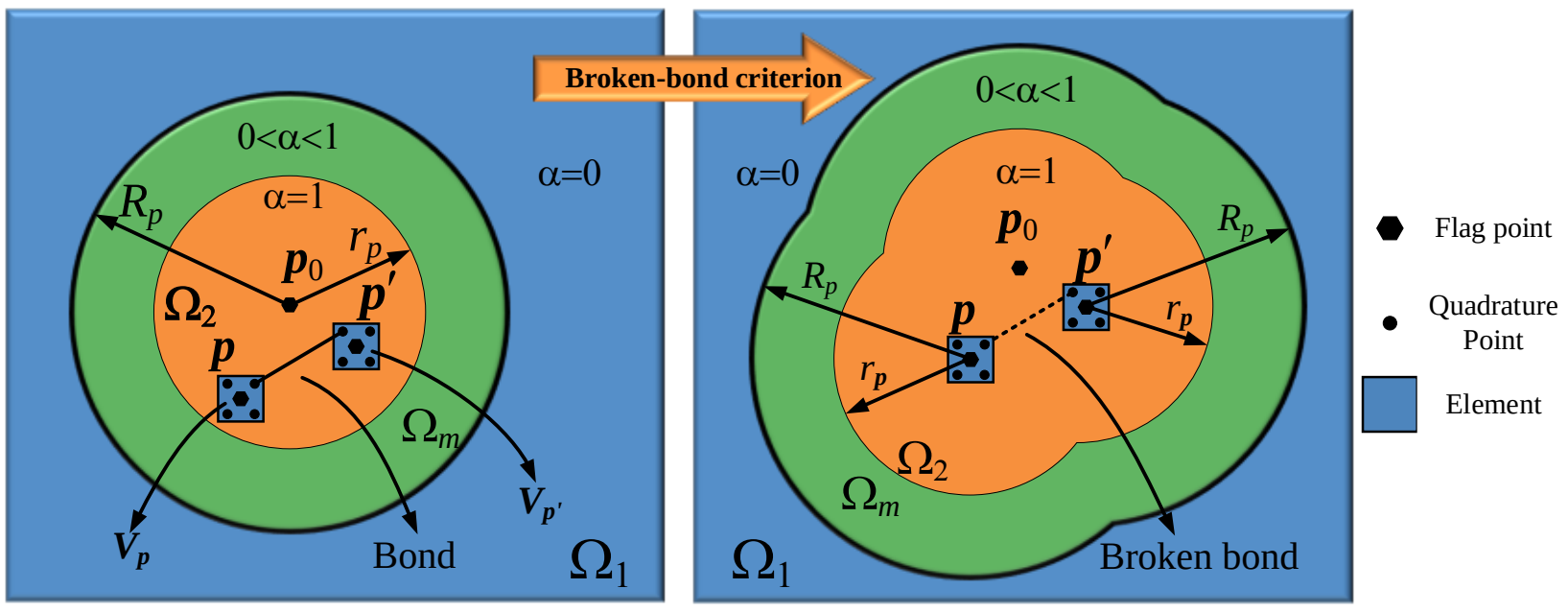

Fig. 3 Flow for adaptive expansion of the PD subdomain driven by the broken bond criterion, where $\boldsymbol{p}_0$ denotes the initial flag point, $\boldsymbol{p}$ and $\boldsymbol{p}'$ are the centroid points of elements $V_p$ and $V_{p'}$, respectively, and the bonds between the quadrature points of the two elements are broken.

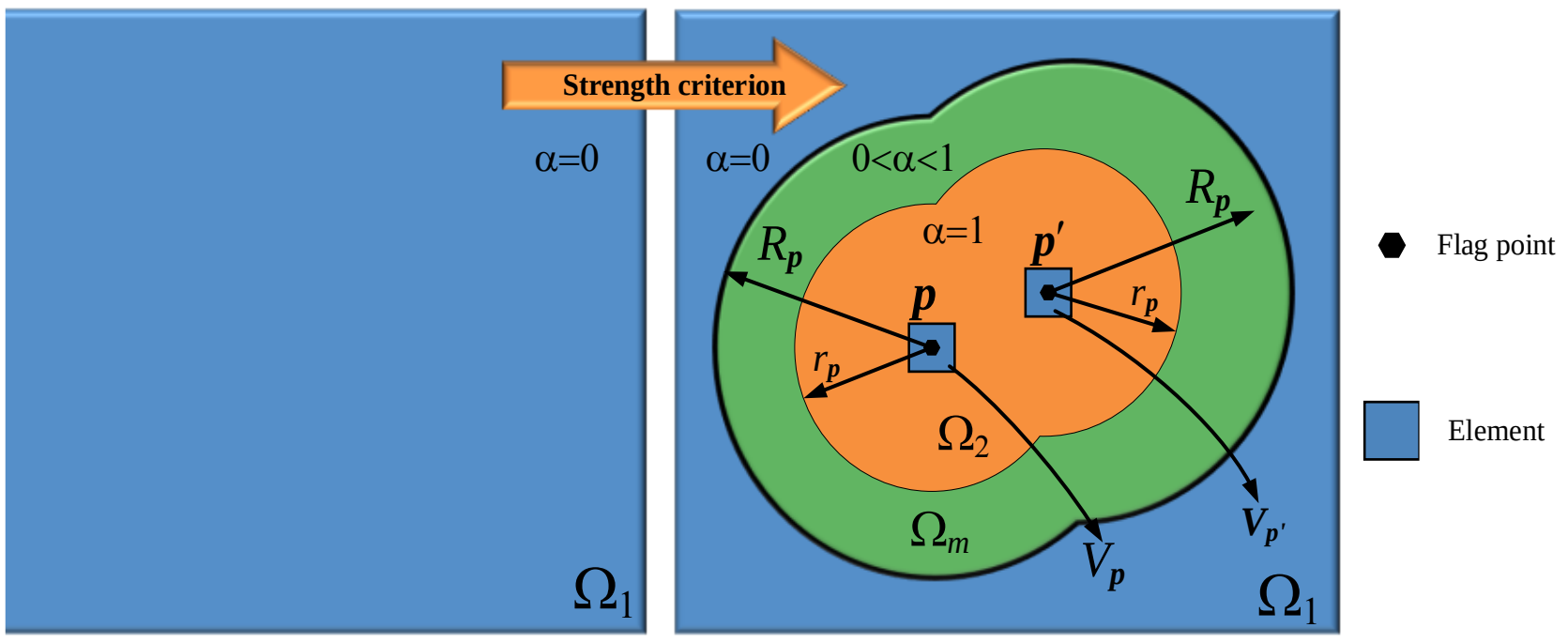

Fig. 4 Flow for adaptive expansion of the PD subdomain driven by the strength criterion, where $\boldsymbol{p}$ and $\boldsymbol{p}'$ are the centroid points of the elements $V_p$ and $V_{p'}$, respectively. The stress states of $\boldsymbol{p}$ and $\boldsymbol{p}'$ reach the critical stress state.

(1) The process for adaptive expansion of the PD subdomain driven by the broken bond criterion, $\mathbb{C}$, is nonempty before the simulation, that is, at time $t=0$. If the initial PD subdomain is arranged as a circular domain centered at the flag point $\boldsymbol{p}_0$ with radius $r_1$, the initial PD model is introduced by assigning the Morphing function $\alpha_{\boldsymbol{p}_0}(\boldsymbol{x})=1$ to the point $\boldsymbol{x}$ within the circular domain $\{\boldsymbol{x} \mid \|\boldsymbol{x}-\boldsymbol{p}_0\| \le r_1\}$, according to the initial condition of the morphing function $\alpha$ described in Eq. (36). Meanwhile, an annular transition

domain with inner radius $r_1$ and outer radius $r_2$ centered at $\boldsymbol{p}_0$ is arranged around the circular domain, where $0 < \alpha_{\boldsymbol{p}_0}(\boldsymbol{x}) < 1$. In contrast, if the process for adaptive expansion of the PD subdomain is driven by the strength criterion, $\mathbb{C}$ is empty at time $t = 0$. Therefore, based on the initial conditions of the Morphing function $\alpha$ described in Eq. (38), $\alpha = 0$ over the entire computational domain $\Omega$. Consequently, the continuum mechanical model is implemented over the entire computational domain.

(2) As the applied load on the structure increases, the elements in the computational domain $\Omega$ begin to be damaged, and the flag-point set $\mathbb{C}$ is determined and updated at each time step according to Eqs. (33) and (35), respectively. Then, the updated value of the Morphing function $\alpha$ can be obtained according to Eqs. (39) and (40).

(3) Since the Morphing function $\alpha = 1$ on the subdomains near the flag points, the PD model appears and takes effect near these flag points.

(4) If the PD bond is stretched beyond the critical value $s_{crit}$, the bond breaks (see Eq. (5)). These broken bonds lead to the redistribution of forces on the structure. An increasing number of elements are damaged around the broken bonds, and the flag-point set $\mathbb{C}$ expands gradually, adaptively activating the PD model over increasingly larger computational subdomains. Subsequently, crack initiation and propagation evolve in the PD subdomain. Fig. 3 and Fig. 4 show the entire process of the emergence and expansion of the PD subdomain driven by the broken bond criterion and the strength criterion, respectively.

## 4. 2 Adaptive expansion radius $r_p$ of a PD subdomain

Here, we consider the proposed model for simulating dynamic crack propagation. As shown in Fig. 5, when a bond breaks at time $t$ or the stress state at a certain point reaches a critical value, a PD subdomain with an inner radius of $r_p$ is applied. Then, at $t + \Delta t$, cracks can initiate and expand in the new PD subdomain. When the crack propagation speed is fast, a sufficiently sized PD subdomain must be applied immediately so the crack can propagate without restriction. Therefore, the inner radius $r_p$ used to expand the new PD subdomain must satisfy the following equation:

$$r_p \geq \boldsymbol{v}_c \Delta t \tag{41}$$

where $\boldsymbol{v}_c$ denotes the crack propagation speed at the current time.

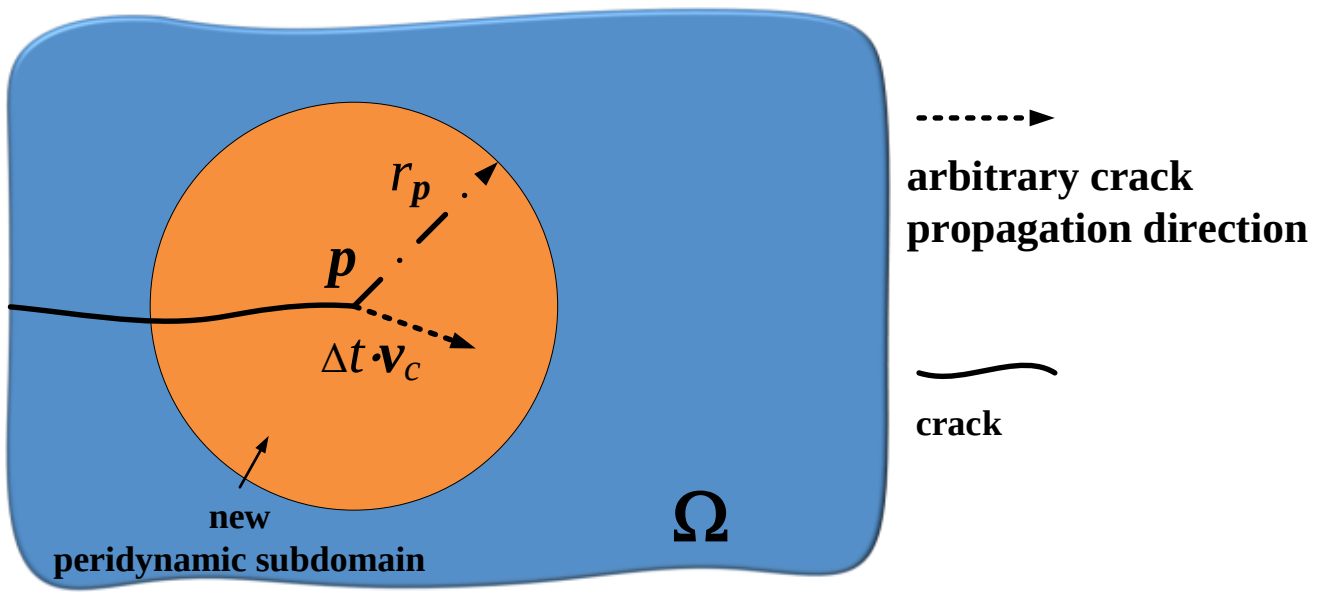

Fig. 5 Crack propagation in the new PD subdomain.

On the other hand, the central difference method, which is a conditional stable algorithm, is used to discretize the time integral in this study. That is, when the central difference method is used to solve a specific problem, the time step $\Delta t$ must be less than some critical value $\Delta t_{cr}$ determined by the equation properties for solving the problem, i.e.,

$$\Delta t \leq \Delta t_{cr} = \frac{2}{\omega_n} = \frac{T_n}{\pi} \tag{42}$$

where $\omega_n$ and $T_n$ are the highest order natural vibration frequency and minimum natural vibration period of the system, respectively. The minimum natural vibration period $T_n$ is always greater than or equal to the minimum natural vibration period of the smallest size element. After the computational domain meshes, it can be calculated using the smallest side length $L$ of the smallest element, where $C = \sqrt{E / \rho}$ is the sound wave propagation speed in the medium, which can be obtained from Eq. (42) as follows:

$$\Delta t_{cr} = L / C \tag{43}$$

where $\Delta t_{cr}$ denotes the time required for the sound waves to pass through the element.

According to the test by Gao [31], the crack propagation speed in solid structures does not exceed the Rayleigh wave speed, and the expression for the Rayleigh wave speed is

$$C_R = \frac{0.862 + 1.14\mu}{1+\mu} C_S \tag{44}$$

where $\mu$ denotes the Poisson's ratio, $C_S = \sqrt{G/\rho}$ denotes the shear wave speed, $G$ denotes the shear modulus, and $\rho$ is the material density. The relationship between $C_S$ and $C$ is given as:

$$C_S = \sqrt{\frac{1}{2(1+\mu)}} C \tag{45}$$

When we substitute Eq. (45) into Eq. (44), the relationship between C and $C_R$ is expressed as follows:

$$C_R = f(\mu) C \tag{46}$$

where $f(\mu) = \frac{0.862 + 1.14\mu}{1+\mu}\sqrt{\frac{1}{2(1+\mu)}} < 1$ always holds; that is, the Rayleigh wave speed $C_R$ is always less than the sound speed $C$. Then, the sufficient condition for Eq. (41) is:

$$r_p \geq C\Delta t_{cr} = L \tag{47}$$

To ensure that the new PD subdomain satisfies the requirements of non-local interactions, the expansion radius $r_p$ of the PD subdomain is generally considered to be greater than the PD horizon $\delta = 3\Delta\bar{x}$, where $\Delta\bar{x}$ denotes the average size of the elements. Therefore, $\Delta\bar{x}$ is often greater than the smallest side length of the smallest element, i.e., $L$. In other words, Eq. (47) always holds; therefore, in each time step, the expansion radius of the PD subdomain $r_p$ cannot affect the crack propagation.

## 4. 3 Numerical algorithm

In this study, the central difference method is used to discretize the time integral in the proposed PD-CCM coupling model. The finite element discretization scheme is shown in Eq. (29), and the nodal acceleration $\ddot{\boldsymbol{u}}(t)$ can be approximated by displacements, i.e.,

$$\ddot{\tilde{\boldsymbol{u}}}(t) = \frac{1}{\Delta t^2}\left[\tilde{\boldsymbol{u}}(t-\Delta t) - 2\tilde{\boldsymbol{u}}(t) + \tilde{\boldsymbol{u}}(t+\Delta t)\right] \tag{48}$$

Substituting the approximate value of the acceleration (i.e., Eq. (48)) into Eq. (29) yields:

$$\frac{\mathbf{M}}{\Delta t^2}\tilde{\boldsymbol{u}}(t+\Delta t) = \boldsymbol{F}(t) - \left(\mathbf{K} - \frac{2\mathbf{M}}{\Delta t^2}\right)\tilde{\boldsymbol{u}}(t) - \frac{\mathbf{M}}{\Delta t^2}\tilde{\boldsymbol{u}}(t-\Delta t) \tag{49}$$

Once the initial two-step displacements $\tilde{\boldsymbol{u}}(t)$ and $\tilde{\boldsymbol{u}}(t-\Delta t)$ are determined, the displacement $\tilde{\boldsymbol{u}}(t+\Delta t)$ at time $t+\Delta t$ can be obtained.

Therefore, we need to consider the initial conditions of Eq. (49). That is , we need to determine the initial displacement $\tilde{\boldsymbol{u}}(0)$ and negative time displacement $\tilde{\boldsymbol{u}}(-\Delta t)$ to calculate the next step displacement $\tilde{\boldsymbol{u}}(\Delta t)$ when $t=0$ . According to the central difference approximation, the nodal velocity can be expressed as

$$\dot{\tilde{\boldsymbol{u}}}(t) = \frac{1}{2\Delta t}\left[-\tilde{\boldsymbol{u}}(t-\Delta t) + \tilde{\boldsymbol{u}}(t+\Delta t)\right] \tag{50}$$

Considering both Eq. (48) and Eq. (50) when $t=0$, the following equation holds:

$$\tilde{\boldsymbol{u}}(-\Delta t) = \tilde{\boldsymbol{u}}(0) - \Delta t\dot{\tilde{\boldsymbol{u}}}(0) + \frac{\Delta t^2}{2}\ddot{\tilde{\boldsymbol{u}}}(0) \tag{51}$$

where $\ddot{\tilde{\boldsymbol{u}}}(0)$ can be calculated using Eq. (29) when $t=0$, i.e.,

$$\ddot{\tilde{\boldsymbol{u}}}(0) = \mathbf{M}^{-1}\left[\boldsymbol{F}(0) - \mathbf{K}\tilde{\boldsymbol{u}}(0)\right] \tag{52}$$

Consequently, the value $\tilde{\boldsymbol{u}}(-\Delta t)$ must be replaced by the initial velocity $\dot{\tilde{\boldsymbol{u}}}(0)$ at $t=0$, and the displacement $\tilde{\boldsymbol{u}}(0)$ and velocity $\dot{\tilde{\boldsymbol{u}}}(0)$ are the initial conditions for Eq. (49). We rewrite Eq. (49) as a linear equation as follows:

$$\hat{\mathbf{M}}\tilde{\boldsymbol{u}}(t+\Delta t) = \boldsymbol{F}(t) - (\mathbf{K} - 2\hat{\mathbf{M}})\tilde{\boldsymbol{u}}(t) - \hat{\mathbf{M}}\tilde{\boldsymbol{u}}(t-\Delta t) \tag{53}$$

where $\hat{\mathbf{M}} = \mathbf{M}/\Delta t^2$.

In the spatial discretization of the proposed model, the continuum mechanical subdomain is meshed by CEs, and the PD subdomain is meshed by DEs. As shown in Fig. 6, when a new PD subdomain is applied, additional nodes must be inserted at the original CE nodal location to convert the CEs into DEs. This increases the degree of freedom of the system at the latter time step, resulting in the displacement vector $\tilde{\boldsymbol{u}}$ and mass matrix $\mathbf{M}$ at different times in Eq. (53) not satisfying the conditions of the same dimensions. To ensure the calculation consistency of the system at the front and back times, we match the displacement vector $\tilde{\boldsymbol{u}}$ and mass matrix $\mathbf{M}$ of the additional nodes with the original node. In the matching principle, the displacement, velocity, and acceleration of nodes at the same position at different times remain consistent, whereas the sum of the node masses at the same position but at different times remains constant.

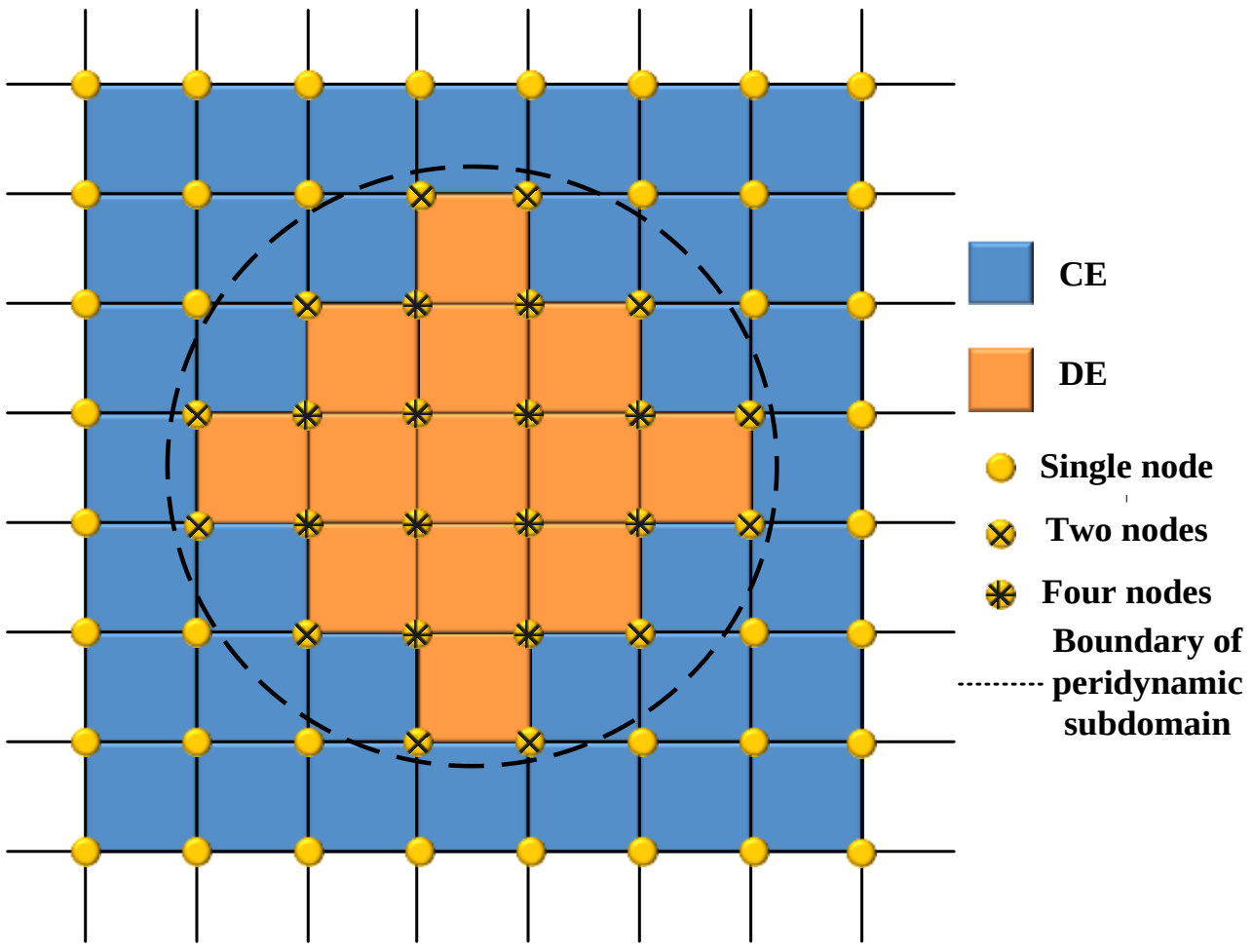

Fig. 6 CEs in the new PD subdomain converted to DEs.

Fig. 7 shows a flowchart of the numerical algorithm, and the specific steps are as follows:

1) Input the initial parameters, such as the time step $\Delta t$ that satisfy $\Delta t < \Delta t_{cr}$, initialize the variable $\kappa_{\alpha}$ that determines whether the PD subdomain expands, and assemble the mass matrix $\mathbf{M}$ and stiffness matrix $\mathbf{K}$ according to the initial mesh data and material parameters.
2) Set the initial boundary conditions, i.e., displacement $\tilde{\boldsymbol{u}}(0)$, velocity $\dot{\tilde{\boldsymbol{u}}}(0)$, and external force $\boldsymbol{F}(0)$.
3) The initial acceleration $\ddot{\tilde{\boldsymbol{u}}}(0)$ is calculated according to Eq. (52). Calculate $\tilde{\boldsymbol{u}}(-\Delta t)$ according to Eq. (51).
4) When the variable $\kappa_{\alpha}$ is true, the mesh data are reconstructed, and the degrees of freedom of the mass matrix $\mathbf{M}$ and nodal displacement vectors $\tilde{\boldsymbol{u}}(t-\Delta t)$ and $\tilde{\boldsymbol{u}}(t)$ in the previous two steps are updated to ensure that the degrees of freedom of each variable in Eq. (53) remains consistent.

5) Calculate $\hat{\mathbf{M}} = \mathbf{M} / \Delta t^2$.

6) The stiffness matrix is updated when new broken bonds are generated. The flag-point set $\mathbb{C}$ is updated by the broken bond criterion or the strength criterion. Then, the Morphing function $\alpha$ on the entire calculation domain $\Omega$ is updated by $\mathbb{C}$ according to Eqs. (39) and (40) to whether a new PD subdomain is applied. If it is expanded, $\kappa_\alpha$ is true. Otherwise, $\kappa_\alpha$ is false.

7) For every time step $(t = 0, \Delta t, 2\Delta t \cdots)$, calculate the effective loading $\boldsymbol{F}_t = \boldsymbol{F}(t) - (\mathbf{K} - 2\hat{\mathbf{M}})\tilde{\boldsymbol{u}}(t) - \hat{\mathbf{M}}\tilde{\boldsymbol{u}}(t - \Delta t)$.

8) Solve the linear equation Eq. (53) to obtain the displacement solution $\tilde{\boldsymbol{u}}(t + \Delta t)$ at time $t + \Delta t$ and update the system displacement field.

9) If needed, calculate the acceleration $\ddot{\tilde{\boldsymbol{u}}}(t)$ and velocity $\dot{\tilde{\boldsymbol{u}}}(t)$ at time $t$ in terms of Eq. (48) and Eq. (50) and perform post-processing.

10) If $t$ is less than the final time, then $t = t + \Delta t$. Return to step 4. Otherwise, end the calculation.

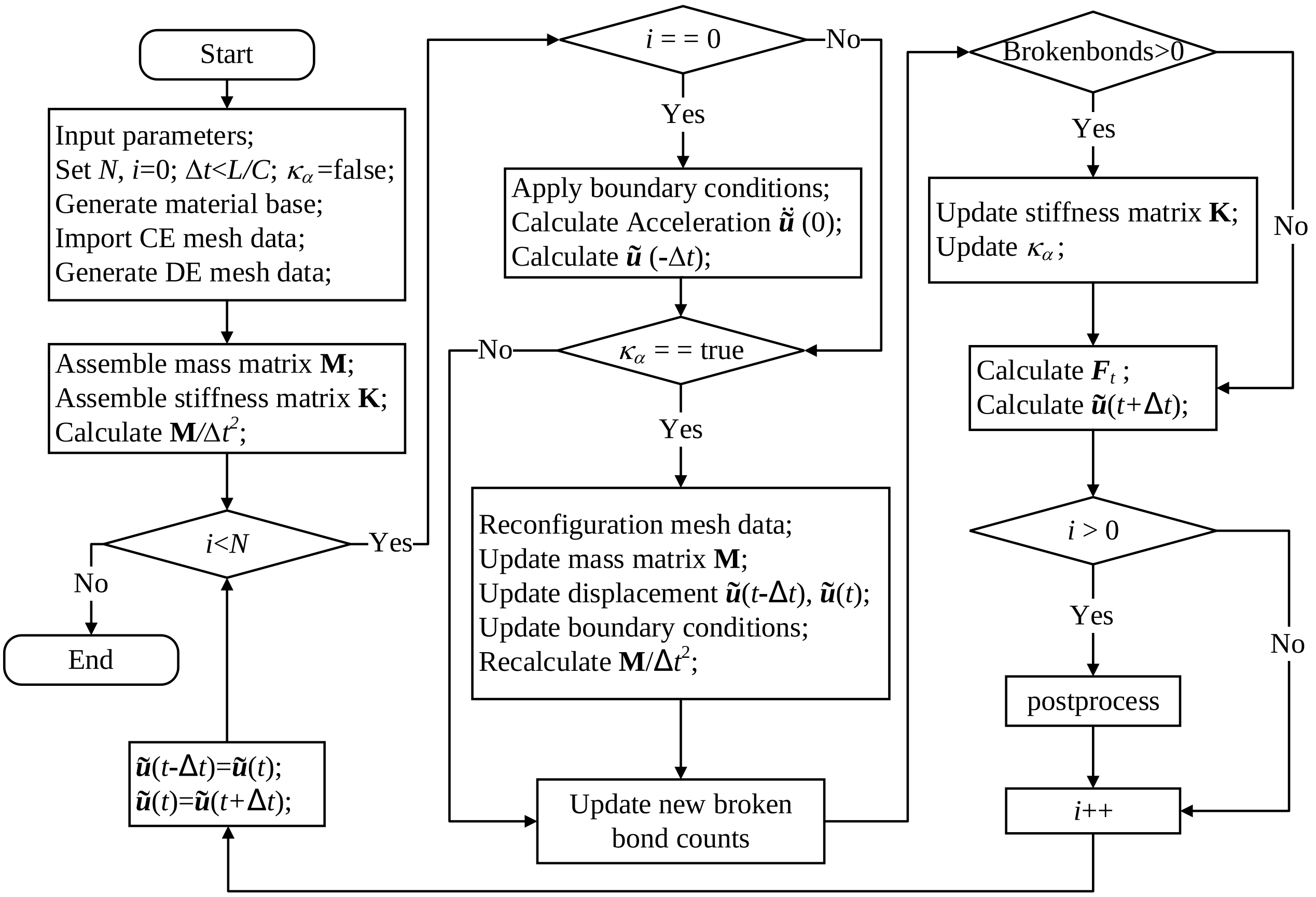

Fig. 7 Flowchart of the numerical algorithm, where $N$ is the total number of time steps in the simulation.

# 5 Numerical examples

In this section, the following three numerical examples are performed to verify the validity, accuracy, and efficiency of the adaptive coupling of the PD-CCM in predicting structural dynamic failure: 1) dynamic crack branching of a pre-notched rectangular plate; 2) dynamic failure of the round disk with a center hole under the explosion load in the hole; and 3) dynamic fracture of a plate with random holes. For dynamic fracture tests, the value of the Poisson ratio $\mu$ has no significant influence on the crack speed and crack path [32]. Therefore, $\mu$ is set as the fixed value of 1/3 in the numerical simulation. The PD horizon radius is set as $\delta = 3\Delta\overline{x}$, where $\Delta x$ is the size of the element, The micro-modulus coefficient is set as $c(\boldsymbol{x},\boldsymbol{\xi}) = \tau^0 e^{-\|\xi\|/l}$, where $\tau^0$ denotes a constant coefficient relating to the Poisson ratio $\mu$ and Young's modulus $E$ [33], and $l$ is the characteristic length, which is chosen as $l = \delta / 15$. All numerical simulations are performed on an 11th Gen Intel (R) Core (TM) i5-11400H CPU computer.

## 5.1 Dynamic crack branching of a pre-notched rectangular plate

The crack branching test of a pre-notched rectangular plate under traction force has been used as a benchmark example in numerous dynamic fracture numerical models [34,35]. This subsection simulates the entire process of crack initiation, propagation, and branching of a pre-notched rectangular plate under the traction force. The simulated crack speed $\boldsymbol{v}_c$ is compared with the pure PD simulation results [36] and experimental results [37], verifying the validity and computational efficiency of PD-CCM adaptive coupling. The geometry and boundary condition are shown in Fig. 8(a), where a $50\ \mathrm{mm}$-long notch is prefabricated from the bottom of the plate along the longitudinal central axis of the plate. The left and right sides of the plate are subjected to a uniform transverse dynamic tensile load $\boldsymbol{\sigma}_x = 14\ \mathrm{MPa}$. The material is soda-lime glass [37], the density of the material is $\rho = 2.44 \times 10^{-9}\ \mathrm{t/mm^3}$, and the Young’s modulus is $E = 72\times10^3$ MPa . The Rayleigh wave speed on the plate is $C_R \approx 3.10\times10^6$ mm/s based on Eq. (44). The energy release rate of the soda-lime glass plate is $G_0 = 1.35\times10^{-4}$ J/mm$^2$ [38]. The size of the mesh element is $\Delta\bar{x} = 0.5$ mm , the progressive time step is set as $\Delta t = 0.04$ μs , and the total simulation time of this example is 38.4 μs .

The initial PD and adjacent transition subdomains are arranged with the pre-notched tip as the center of the circle with radii of 6 mm and 12 mm , respectively, to implement the adaptive coupling of the PD-CCM driven by the broken bond criterion, as shown in Fig. 8(b). Meanwhile, the fixed coupling of the PD-CCM shown in Fig. 8(c) is generated as the reference model. In this fixed coupling of PD-CCM, the longitudinal symmetry axis of the plate is taken as the center line, and the initial PD and adjacent transition subdomains are set with widths of 16 mm and 19 mm , respectively.

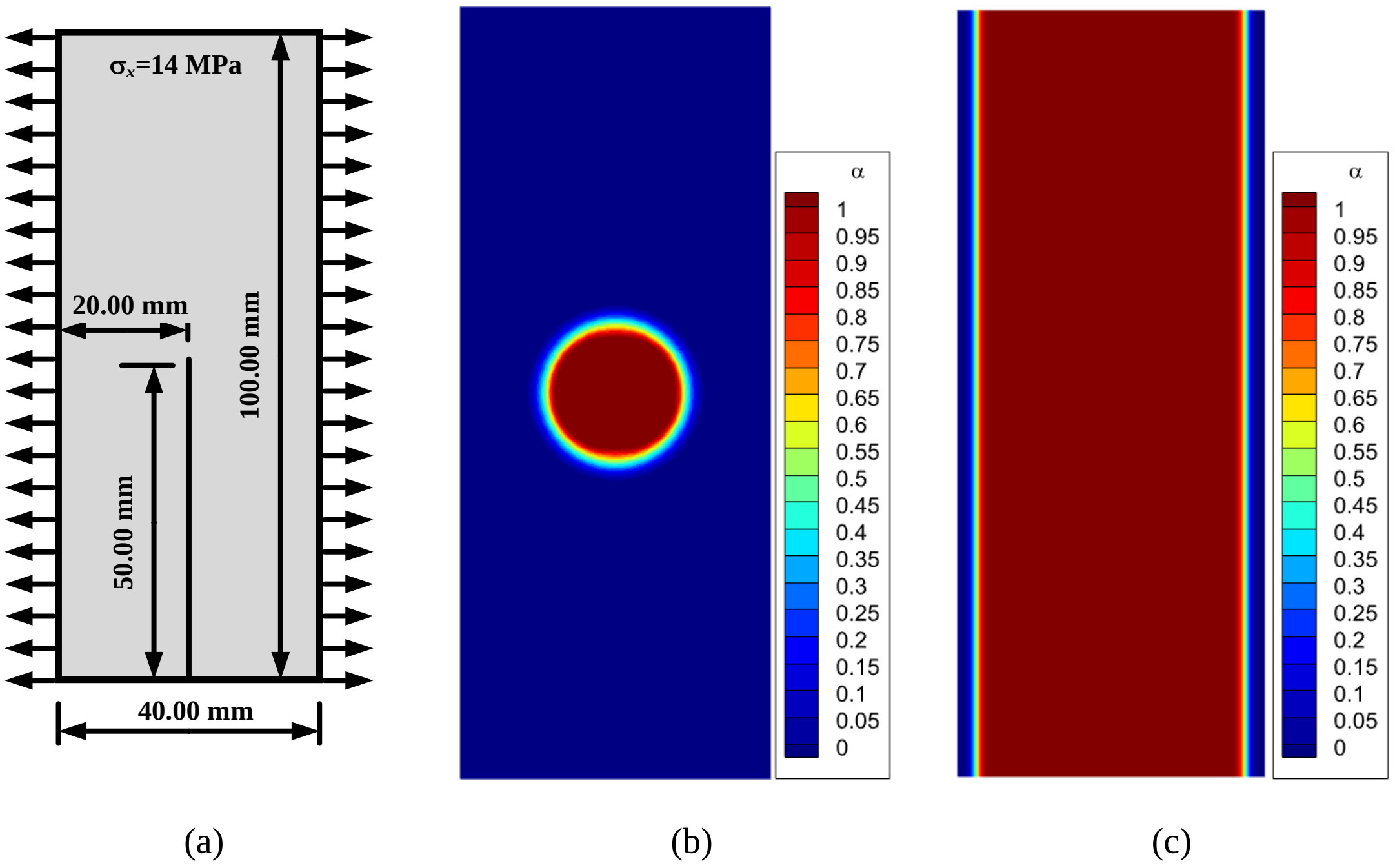

Fig. 8 Initial setup of a rectangular plate with a pre-notch starting from the bottom along the longitudinal central axis under dynamic tensile load: (a) geometry and boundary condition; (b) contour of the initial Morphing function for the adaptive coupling of the PD-CCM; (c) contour of the initial Morphing function for the fixed coupling of the PD-CCM.

Fig. 9 illustrates the evolution process of crack initiation, propagation, and branching of the pre-notched rectangular plate, where the white grids depict the range of PD subdomains. We observe that the crack paths predicted by the two models are consistent by comparing the subgraphs in Fig. 9(a) and (b). The crack first appeared at the pre-notched tip at time 10 μs. Propagation begins along the longitudinal direction, and crack branching occurs at the main crack tip at time 20.8 μs. Then, the two branched sub-cracks propagate upward for a distance within the longitudinal central axis of the plate, and after deflecting to the longitudinal central axis, the propagation direction continues to propagate upward obliquely at time 33.6 μs. Finally, the two branched sub-cracks continue to propagate upward after the deflection of the propagation direction, and the entire plate is completely destroyed at time 38.4 μs. The above results indicate that the adaptive coupling of the PD-CCM can effectively capture the entire process of crack initiation, propagation, and branching [25].

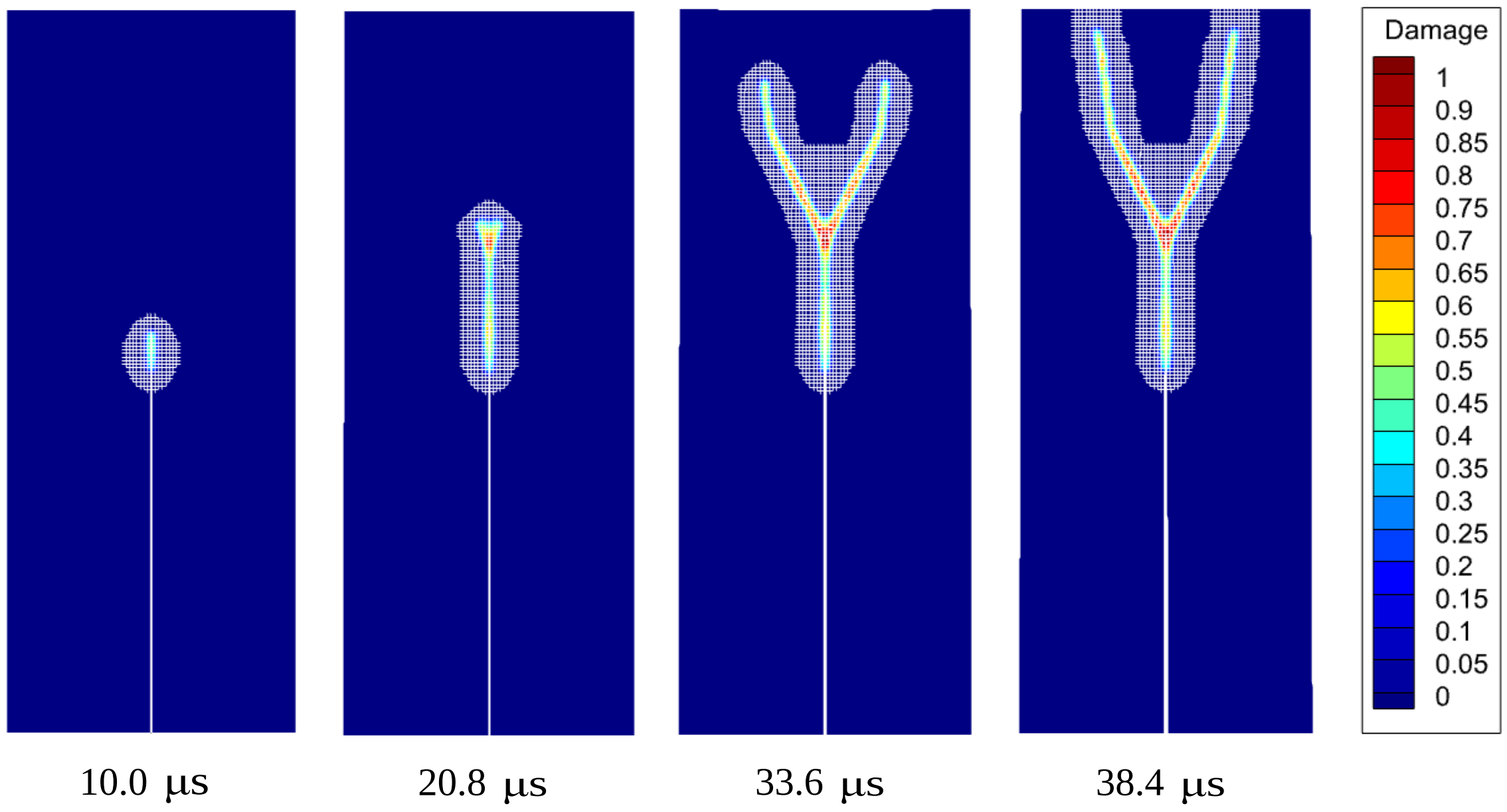

(a)

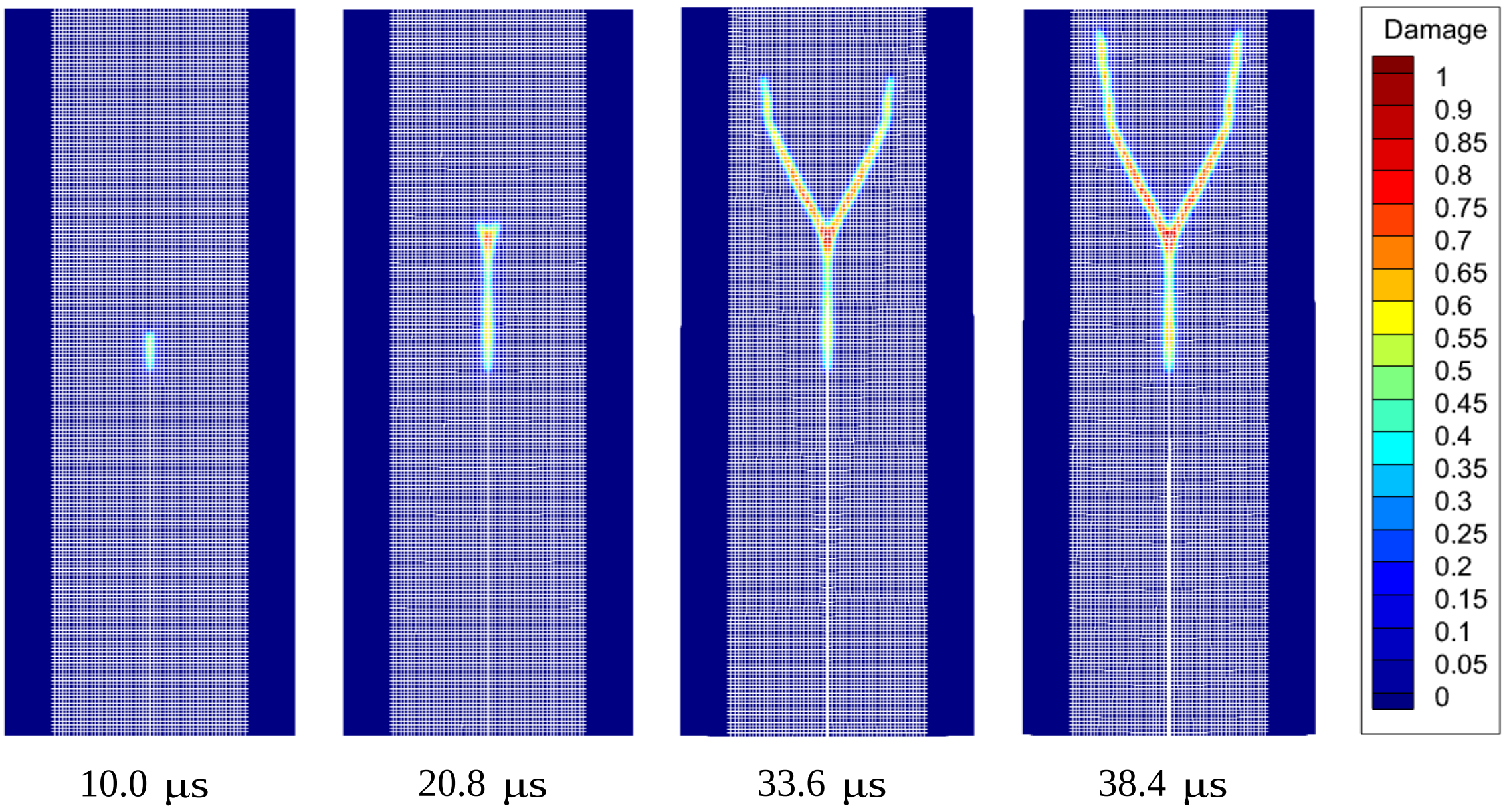

(b)

Fig. 9 Damage contour of the pre-notched rectangular plate under dynamic tensile load at different times: (a) simulation process of the adaptive coupling of the PD-CCM and (b) simulation process of the fixed coupling of the PD-CCM.

Fig. 10 compares the calculation time at each time step of the PD-CCM adaptive coupling with the PD-CCM fixed coupling. For the former model, the calculation time is almost zero when the time step is between 0 and 100. This is because no cracks on the plate are initiated at this stage, and the calculation time of the program is mainly consumed in the displacement computation process using the central difference method without updating the stiffness and mass matrices. When the time step is between 100 and 960, the calculation time increases with the increased time step. This is because steady crack propagation promotes the expansion of the PD subdomains, which increases the degrees of freedom in the coupled system. Alternatively, in terms of the fixed coupling of the PD-CCM, when the time step is between 0 and 130, the calculation time of each time step is short. However, it is slightly higher than the adaptive coupling of the PD-CCM. The time is longer than that of the adaptive model because the PD subdomains of the fixed coupling of the PD-CCM are larger, and the overall degree of freedom of the system is higher. When the time step reaches 130, the calculation time increases sharply and then remains constant (approximately 9 s per time step) as the time step increases. This is because the stiffness matrix is updated at each step with the appearance of bond breakage.

Fig. 10 illustrates that the crack initiation time of the adaptive coupling of the PD-CCM is slightly earlier than the fixed coupling of the PD-CCM. This may be due to the boundary effect of PD (the weakening of the equivalent properties of materials near the boundary), making the fixed coupling of the PD-CCM with a large PD subdomain more flexible than the adaptive model and, thus, harder to crack. This phenomenon is also observed in Fig. 11, where the crack initiation time predicted by the adaptive coupling of the PD-CCM is earlier than the pure PD model in the literature [36]. Regarding the total calculation time, the simulation process of the fixed coupling of the PD-CCM takes approximately 7843.170 s, and that of the adaptive coupling of the PD-CCM takes approximately 3156.555 s.

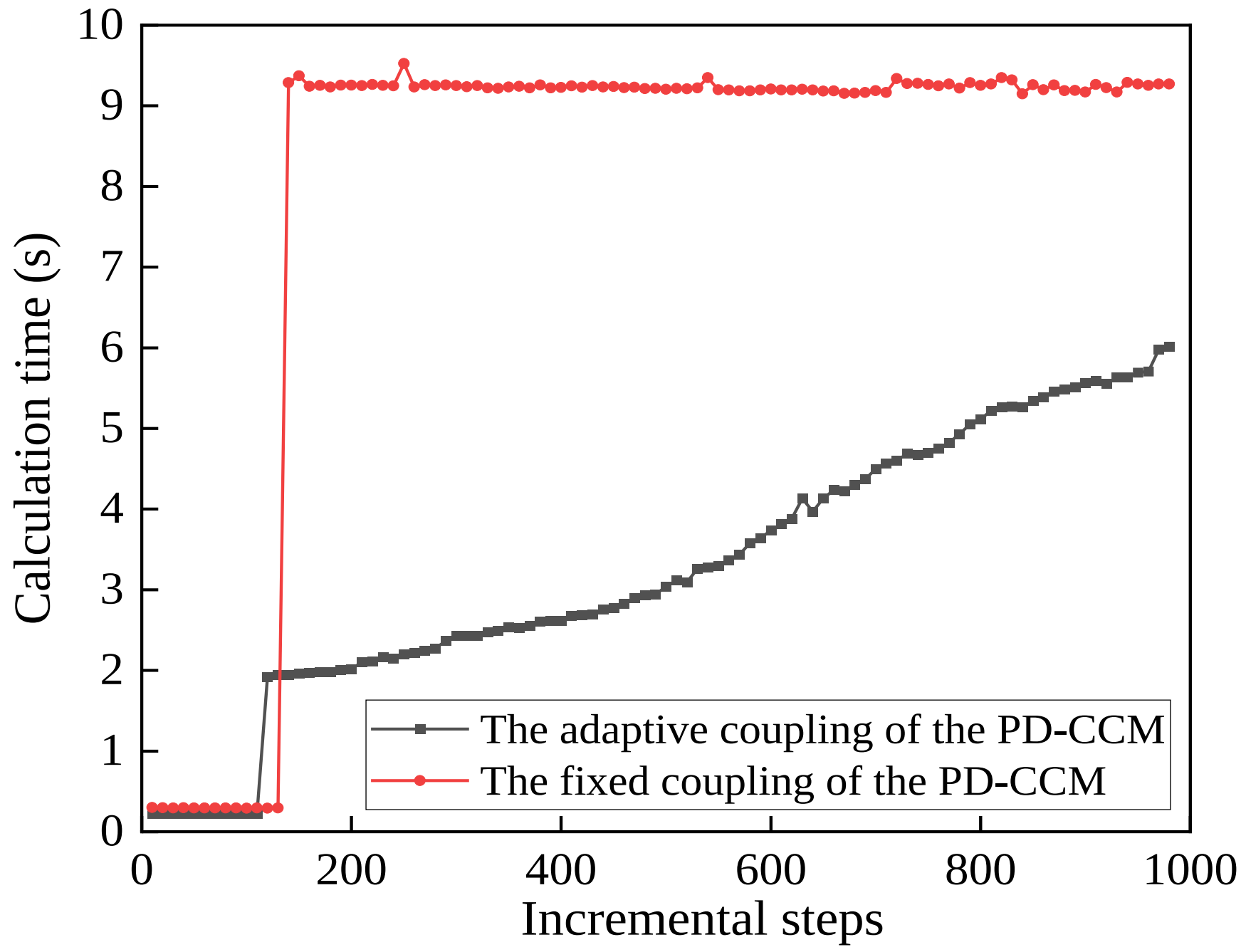

Fig. 10 Comparison of calculation times at each time step of the adaptive coupling of the PD-CCM and the fixed coupling of the PD-CCM.

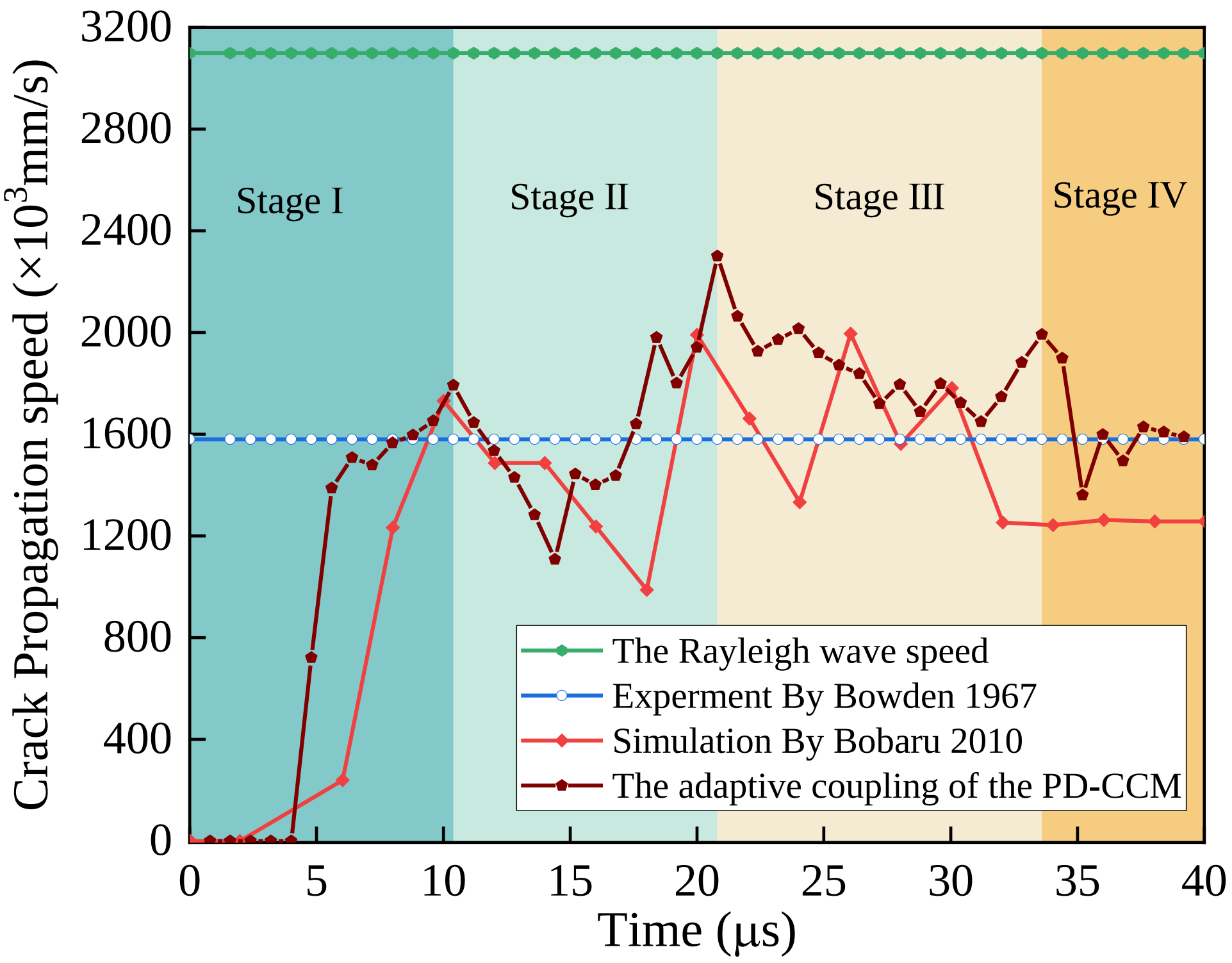

Fig. 11 Crack speed $\boldsymbol{v}_c$ under dynamic tensile load.

Fig. 11 illustrates the comparison between the crack speed $\boldsymbol{v}_c$ predicted by the adaptive coupling of the PD-CCM, the pure PD model [36], and the experimental results [37]. The developmental course of $\boldsymbol{v}_c$ with time step increase predicted by the adaptive coupling of the PD-CCM model can be divided into four stages. The damage contour corresponding to the critical time and the last time used to divide the two adjacent stages is presented in Fig. 9(a). In the first stage, the crack in the rectangular plate appears, and $\boldsymbol{v}_c$ increases rapidly from $0\ \text{mm/s}$ to $1.80\times10^6\ \text{mm/s}$. In the second stage, $\boldsymbol{v}_c$ gradually decreases to a minimum, and the cracks begin branching. Once the cracks branch, $\boldsymbol{v}_c$ rapidly increases to $2.30\times10^6\ \text{mm/s}$. In the third stage, $\boldsymbol{v}_c$ gradually reaches its highest speed until the crack propagation direction is deflected. When the propagation direction of the branched sub-cracks is symmetrically deflected towards the central axis, $\boldsymbol{v}_c$ accelerates again and increases to $1.99\times10^6\ \text{mm/s}$. In the fourth stage, the cracks gradually propagate to the top of the plate. Meanwhile, $\boldsymbol{v}_c$ is limited, decreasing gradually and eventually stabilizing around the results measured in the experiment [37].

Fig. 11 also shows that the evolution trend of $\boldsymbol{v}_c$ simulated by the adaptive coupling of the PD-CCM and the pure PD model [36] is consistent. During the entire evolution process of the crack, the crack speed is mainly concentrated near the speed measured by the experiment [37], and it never exceeds the Rayleigh wave speed $C_R$.

## 5. 2 Dynamic failure of a round disk with a center hole under the explosion load in the hole

This subsection describes the dynamic failure process of a round disk with a center hole under an explosion load in the hole to verify the validity of the proposed model. The experiment sample referencing the rock blasting experiment in [39] is shown in Fig. 12(a). The two-dimensional geometric model is established as shown in Fig. 12(b). A hole with a diameter of $6.45\ \text{mm}$ is set in the center of a disk with a diameter of $144\ \text{mm}$. The material of the disk is called Barre granite rock [39], the density of the material is $\rho = 2.70\times10^{-9}\ \text{t/mm}^3$, the Young's modulus is $E = 72\times10^3\ \text{MPa}$, and the energy release rate is $G_0 = 2.217\times10^{-2}\ \text{J/mm}^2$. The calculation domain is meshed by the unstructured elements, in which the average size is $\Delta\bar{x} \approx 0.5\ \text{mm}$. The progressive time step is set as $\Delta t = 0.05\ \mu\text{s}$, and the total simulation time of this example is $50\ \mu\text{s}$.

The adaptive coupling of the PD-CCM driven by the broken bond criterion is implemented, as shown in Fig. 12(c). The initial PD and adjacent transition subdomains are arranged, with the center of the hole as the center of the circle, having radii of $6\ \text{mm}$ and $12\ \text{mm}$, respectively. The outer boundary of the disk is a free boundary. The explosion load is imposed on the boundary of the hole, and the time-history curve of the explosion load is shown in Fig. 12(d). In this study, the explosion load is described using the formula proposed by Trivino [40]:

$$\boldsymbol{P}(t) = \boldsymbol{P}_0 \cdot P_u(t) \cdot P_d(t)$$

$$P_u(t) = e^{-\left[e/(2g)\cdot m_u\cdot(t-t_u)\right]^{2g}},\ P_d(t) = e^{-\left[\sqrt{2e}/2\cdot m_d\cdot(t-t_d)\right]^2}$$

where $m_u$, $m_d$, $g$, $t_u$, and $t_d$ are constant parameters. $g$, $t_u$, and $t_d$ are functions of the maximum rise rate $m_u$ and the maximum decay rate $m_d$ :

$$g = \mathrm{round}(\sqrt{2e} \cdot m_u / m_d)$$

$$t_u = \frac{[-\ln(\alpha_1)]^{1/2g}}{e/(2g) \cdot m_u}, \; t_d = \frac{[-\ln(\alpha_1)]^{1/2g} - [-\ln(1-\alpha_2)]^{1/2g}}{e/(2g) \cdot m_u}$$

where the round function denotes the nearest integer. In this study, we select $\alpha_1 = 1 \times 10^{-7}$ and $\alpha_2 = 1 \times 10^{-3}$. $t_d$ is the time of peak load. With the increase of $m_u$ and $m_d$ , the maximum rise and decay rates of the time-history curve of the explosion load increase. The explosion load parameters are $\boldsymbol{P}_0 = 500$ MPa , $m_u = 9 \times 10^5$ , $m_d = 1 \times 10^5$, $t_d = 3.8$ μs , and $g = 21$.

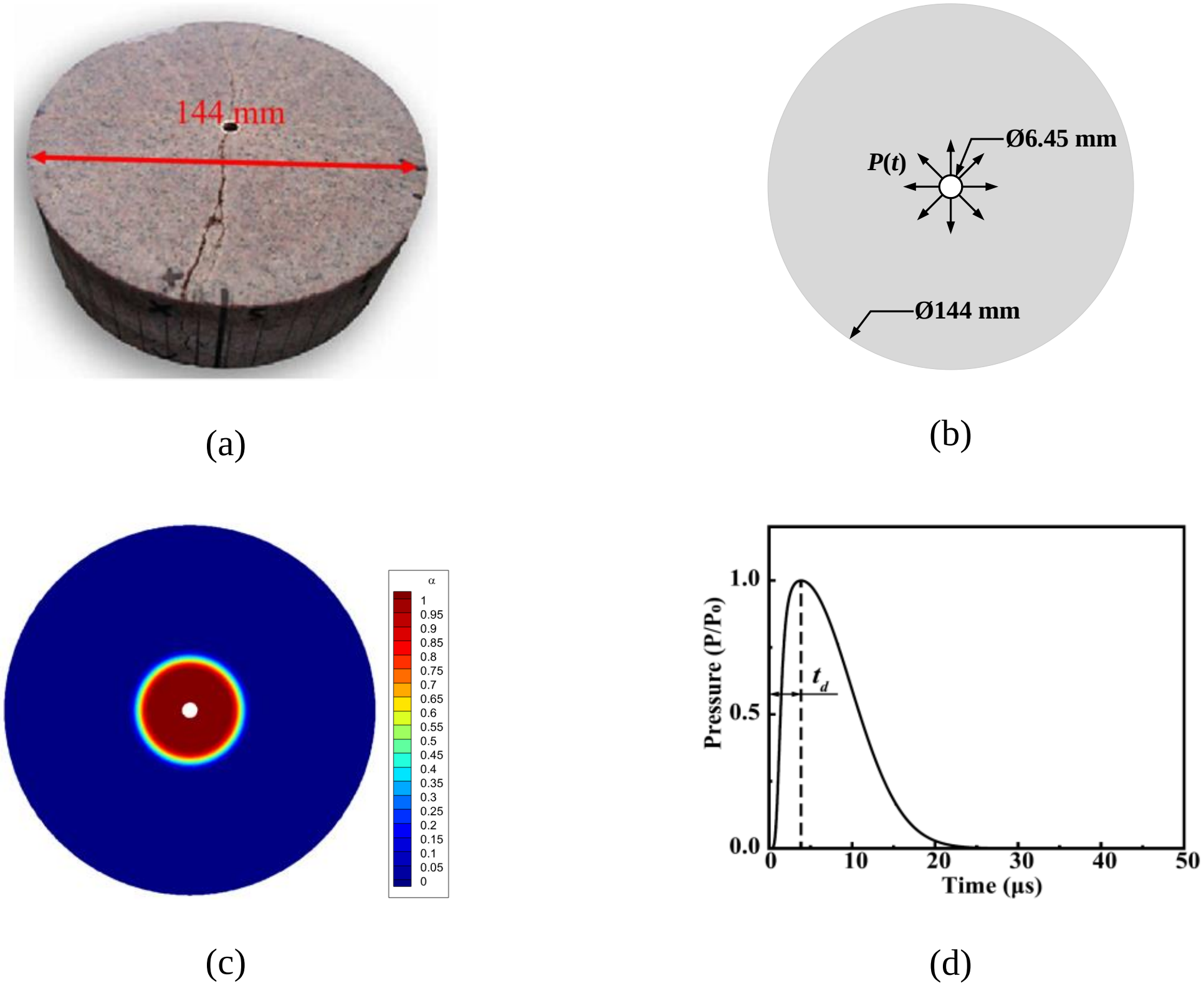

(a) (b) (c) (d)

Fig. 12: The initial setup of the disk with a center hole under the explosion load in the hole: (a) the photo of the test sample; (b) the geometry and the boundary condition; (c) the contour of the initial Morphing function for the adaptive coupling of the PD-CCM; (d) the time-history curve of the explosion load, where $t_d$ denotes the peak time of the explosion load, $\boldsymbol{P}_0$ denotes the peak pressure of the explosion load, $\boldsymbol{P}$ denotes the corresponding explosion load at different times, and $\boldsymbol{P}/\boldsymbol{P}_0$ is the normalized explosion load.

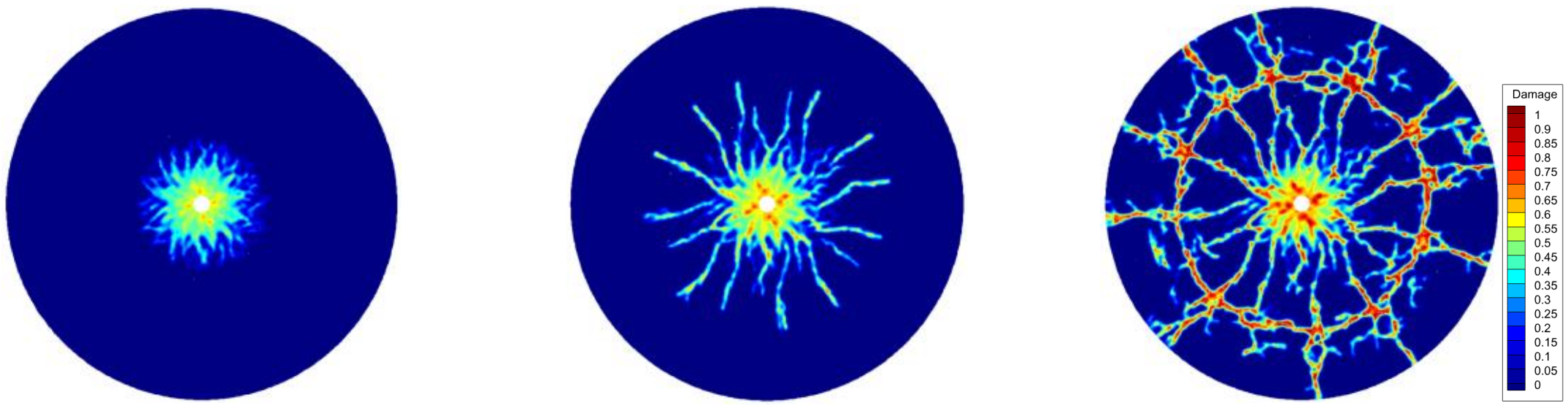

(a) 7 μs (b) 20 μs (c) 50 μs

Fig. 13 Damage contours of the disk under an explosion load at different times.

Fig. 13 illustrates the dynamic failure process of the disk simulated by the adaptive coupling of the PD-CCM under the explosion load. When the time of explosion load imposed on the hole of the disk is $7\,\mu s$, a dense damage domain appears around the hole, and the edge of the damage domain contains several small cracks (see Fig. 13(a)). A similar phenomenon can also be observed in the experiment [41]. This is because the explosion load causes a violent explosion shock wave with significantly greater strength, which exceeds the dynamic compressive strength of the rock. Consequently, the rock around the blasting hole is excessively crushed. When the time of explosion load is $20\,\mu s$, more than ten tortuous radial cracks appear outside the dense damage domain and expand toward the free boundary of the disk, as shown in Fig. 13(b). From the experimental observation [41], this phenomenon is because when the explosion shock wave is attenuated into a compression wave, the tensile and shear waves are generated behind the compression wave, continuously generating new and more robust radial cracks inside the rock. When the time of explosion load is $50\,\mu s$, a ring crack band is generated on the inner side near the free boundary of the disk. Meanwhile, 11 radial cracks continue propagating to the free boundary through the ring crack band, while the remaining cracks are blocked and stop propagating (as shown in Fig. 13(c)). Branching and tangential propagation of cracks appear between the crack band and the free boundary. From the experimental observation [41], this is because when the compression wave reaches the free boundary of the rock, the incident compression wave is transformed into a tensile wave and reflected back to the interior of the rock, and complex cracks such as spalling cracks appear near the free boundary.

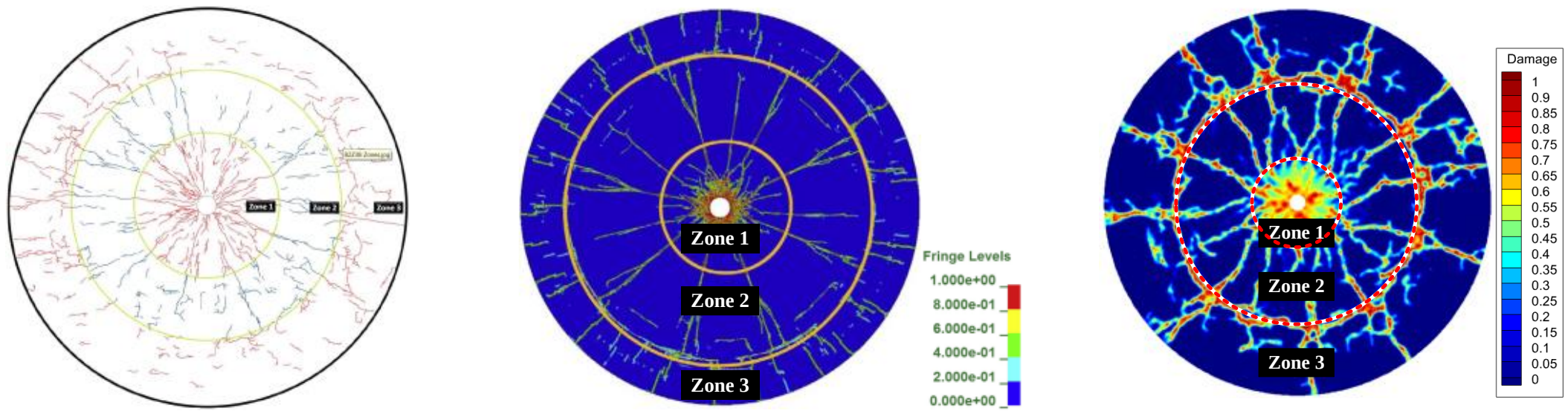

(a) (b) (c)

Fig. 14 Final crack morphology of a disk with a center hole under an explosion load in the hole: (a) the experimental results [37]; (b) the simulation results by LS-DYNA [40]; and (c) the simulation results by the adaptive coupling of the PD-CCM.

Fig. 14 illustrates the final crack morphology of the disk with a center hole under an explosion load in the hole simulated by the experiment [37], LS-DYNA [40], and the adaptive coupling of the PD-CCM. By comparing each subgraph in Fig. 14, we find that the final crack morphology shown in the three subgraphs can be divided into the following three zones: zone 1) the compressive failure zone under the explosion shock wave; and zone 2) the tensile shear failure zone under the combined actions of the tensile and shear waves; zone 3) the spalling failure zone under the action of the reflected stretch wave. Compared with the LS-DYNA results shown in Fig. 14(b), the crack morphology simulated by the proposed model shown in Fig. 14(c) is closer to the crack morphology observed in the experiment shown in Fig. 14(a). The irregularity of the crack path in the experiment and the details of the crack morphology near the free boundary are well presented in the proposed model.

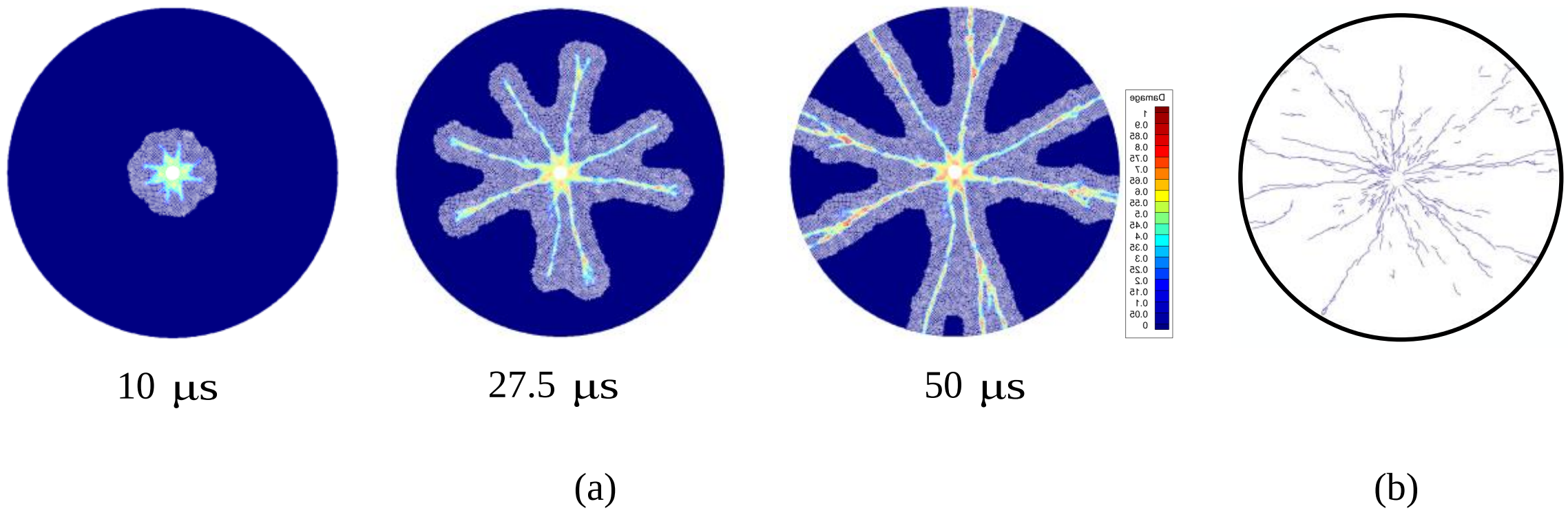

Fig. 15 Damage contour of the disk with a center hole under an explosion load in the hole: (a) the damage contour simulated by the adaptive coupling of the PD-CCM at different times; (b) the experimental results [39].

This subsection simulates the dynamic failure process of the disk at peak load time $t_d = 10\ \mu s$ to demonstrate the ability of the proposed model to efficiently drive the expansion of the PD subdomains. Fig. 15(a) shows the dynamic failure process of the disk at three moments, $10\ \mu s$, $27.5\ \mu s$, and $50\ \mu s$, where the white-grid domain denotes the calculation domain of the PD model. Fig. 15(b) shows the crack morphology finally observed in the experiment [39]. The white-grid domain, that is, the PD subdomains, gradually expands with crack propagation and covers the crack with a small domain at all times (as shown in Fig. 15). The white-grid domain adaptively branches and merges as the cracks branch and merge. The predicted crack morphology at time $50\ \mu s$ in Fig. 15(a) is consistent with the experimental observation in Fig. 15(b). Thus, this further proves that the adaptive coupling of the PD-CCM can effectively predict the dynamic fracture process while minimizing the use of the PD subdomain to reduce the computational cost as much as possible. Therefore, the proposed adaptive coupling of the PD-CCM has considerable practical value.

## 5. 3 Dynamic fracture of a plate with random pores

The above two examples are simulated based on the adaptive coupling of the PD-CCM driven by the broken bond criterion. In this subsection, an adaptive coupling of the PD-CCM driven by strength criteria is applied to simulate the fracture process under dynamic tensile load. The strength criterion is employed to avoid the problem of setting the initial PD subdomain in a plate with several random pores because the crack initiation location cannot be estimated in advance for this plate type. Subsequently, the simulation results are compared with the experimental [43] and simulation results [43][44], verifying the validity of the adaptive coupling of the PD-CCM driven by the strength criterion. The geometry and the boundary condition of the plate with random pores are shown in Fig. 16, in which 31 circular pores with the same radius of $3.175$ mm are randomly generated on the plate, and the distribution positions of the pores are consistent with those in the experiment [43]. The plate material is epoxy, with density $\rho = 1.10\times10^{-9}\ \mathrm{t/mm^3}$, and the Young's modulus is $E = 3.26\times10^{3}\ \mathrm{MPa}$. The experiment [45] reveals that the axial tensile strength of epoxy is $\sigma_{crit} = 62.86\ \mathrm{MPa}$. The critical stretch of the bond is set as $s_{crit} = 0.03$. The calculation domain is meshed by unstructured elements with average size $\Delta\bar{x} \approx 0.5\ \mathrm{mm}$. The progressive time step is set as $\Delta t = 0.01\ \mu\mathrm{s}$, and the total simulation time of this example is $20\ \mu\mathrm{s}$. The dynamic tensile load is imposed on the upper boundary of the plate, and its specific expression is as follows:

$$\sigma_y(t) = \begin{cases} \sigma_0\, t/t_0 & \text{for} \quad t \le t_0 \\ \sigma_0 & \text{for} \quad t > t_0 \end{cases} \tag{54}$$

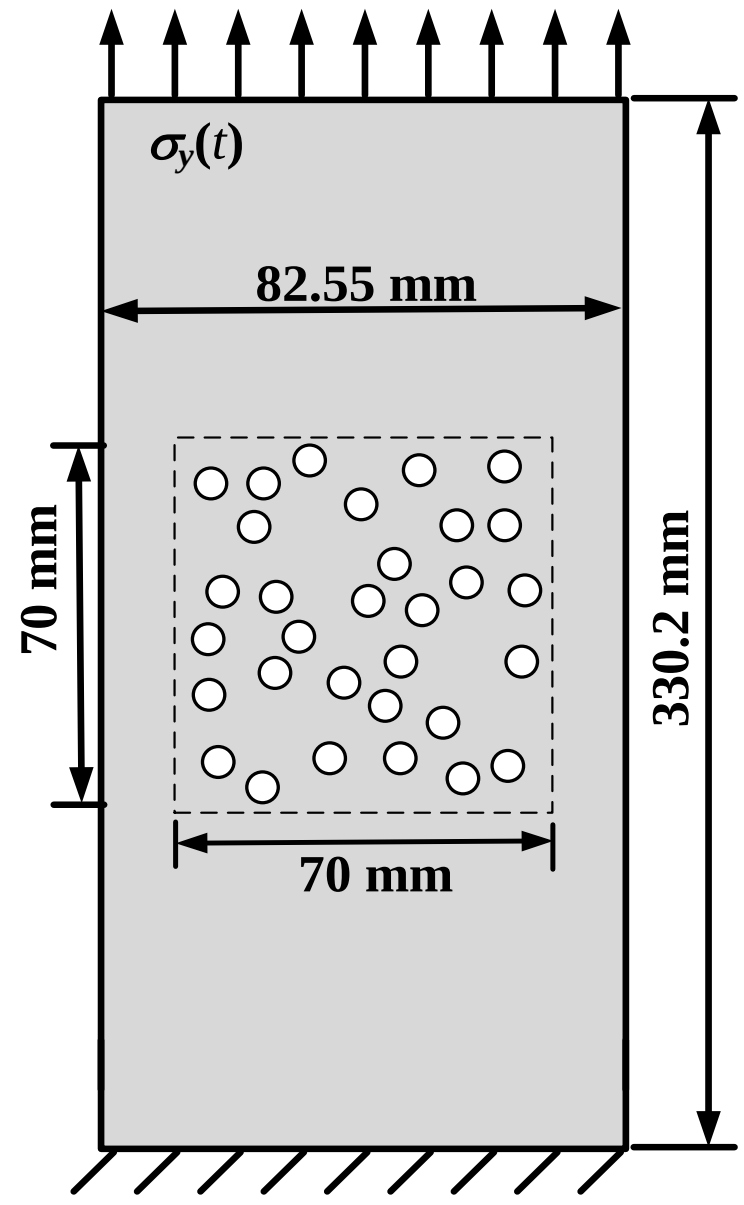

Fig. 16 Geometry and boundary conditions of the plate with random pores.

Fig. 17 illustrates the final crack morphologies on the plate under the load conditions of $\sigma_0 = 12$ MPa and $t_0 = 5$ μm, in which Fig. 17(a-d) displays the experimental results [42], the simulation results of the proposed model, and the simulation results in the literature [44,43]. The experimental results [44] in Fig. 17(a) show that the crack path is dependent on the pore locations. The main crack begins at pore A and propagates to the right. When passing through pore B, the main crack branches to form two sub-cracks. The upper sub-crack continuously passes through the pore near the crack and then penetrates the plate to the right through pore E. On the other hand, the lower sub-crack first passes through pore C, and then secondary branching occurs after pore C. Next, the upper-branched crack propagates to pore D, and the lower-branched crack continues to propagate to the right through pore F. Fig. 17(b) illustrates the simulation results of the proposed model, which is consistent with the experimental results. The main crack starts at pore A, and the crack branching occurs near pore B and then pore C. Then, the upper branched crack propagates through the plate from pore C through pore D to pore E, and the lower branched crack propagates diagonally from pore C to pore F. Fig. 17(c) illustrates the simulation results in the literature [44]. Although the process of crack branching is similar to the experimental results, the exact locations of crack initiation and branching are quite different from the experimental observations. Fig. 17(d) shows the simulation results reported in the literature [43]. In this figure, the crack initiation location is higher than that of pore A in the experimental results, and the crack branching is not well simulated. In summary, the final crack morphology predicted by the proposed model is most similar to the experimental results.

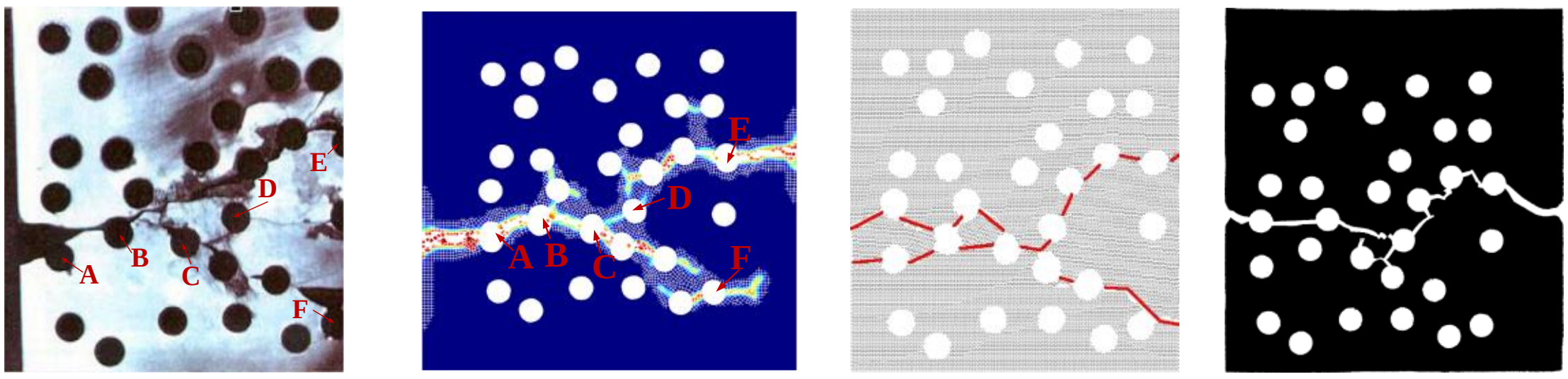

(a) (b) (c) (d)

Fig. 17 Final crack morphologies in the plate: (a) the experimental results [44]; (b) the simulation results of the adaptive coupling of the PD-CCM; (c) the simulation results from the literature [44]; and (d) the simulation results from the literature [43].

Fig. 18 illustrates the final crack morphologies when the maximum tensile load is 10, 15, 20, and 25 MPa, respectively. As the maximum tensile load increases, the initial position of the main crack gradually moves upward. When the maximum tensile load $\sigma_0 = 10$ MPa, the main crack first passes through a random pore located near the bottom of the plate. Fig. 18(b) shows that when $\sigma_0 = 15$ MPa, the initiation of the main crack occurs in the middle of the plate, and crack branching occurs when the crack propagates to the right. Fig. 18(c) shows that when $\sigma_0 = 20$ MPa, the position of the main crack moves further upward. The crack first passes through the top three pores near the left side and then propagates diagonally to the lower right. Fig. 18(d) shows that when $\sigma_0 = 25$ MPa, the main crack penetrated all the pores at the top of the plate. Additionally, the white-grid domain is a PD subdomain (see all subfigures in Fig. 18). The PD subdomain occupies only a small part of the entire computational domain until the crack finally passes through the plate. Therefore, the computational consumption of the adaptive coupling model proposed in this study is greatly reduced compared to using the PD model in the entire computational domain.

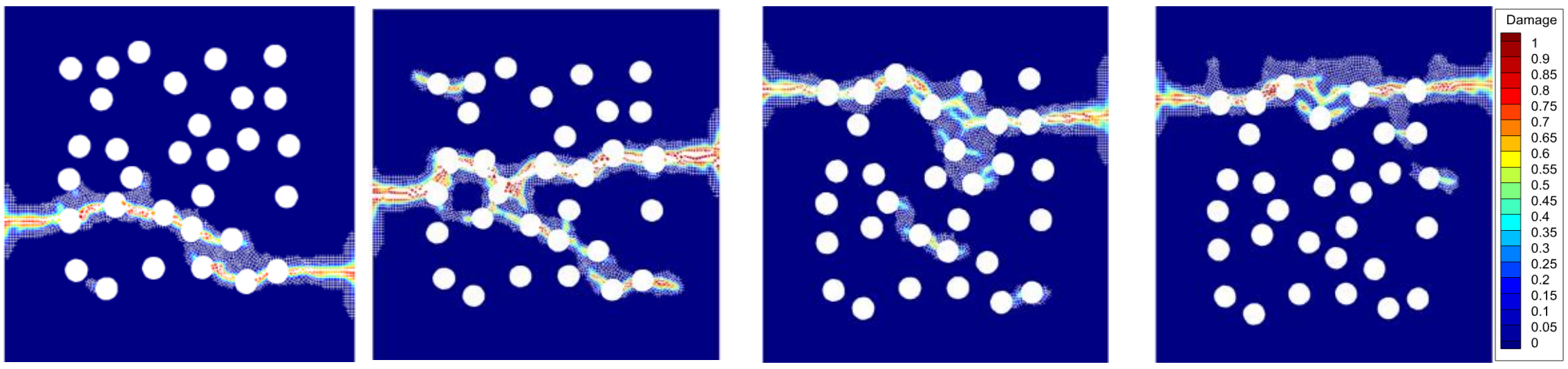

(a) $\sigma_0 = 10$ MPa (b) $\sigma_0 = 15$ MPa (c) $\sigma_0 = 20$ MPa (d) $\sigma_0 = 25$ MPa

Fig. 18 Final crack morphology on the plate under different maximum tensile loads

Fig. 19 illustrates a summary diagram integrating all the crack paths shown in Fig. 18 and those observed in the experiments [43]. The cracks are mainly concentrated in two domains: 1) around the two rows of the pores at the top; 2) around a row of pores in the middle of the plate; and also around a row of pores oblique to the lower right of the plate due to crack branching. In summary, the adaptive coupling of the PD-CCM driven by the strength criterion accurately predicts the location of crack initiation and provides a more reasonable crack path. The proposed model can also minimize computational costs and improve computational efficiency.

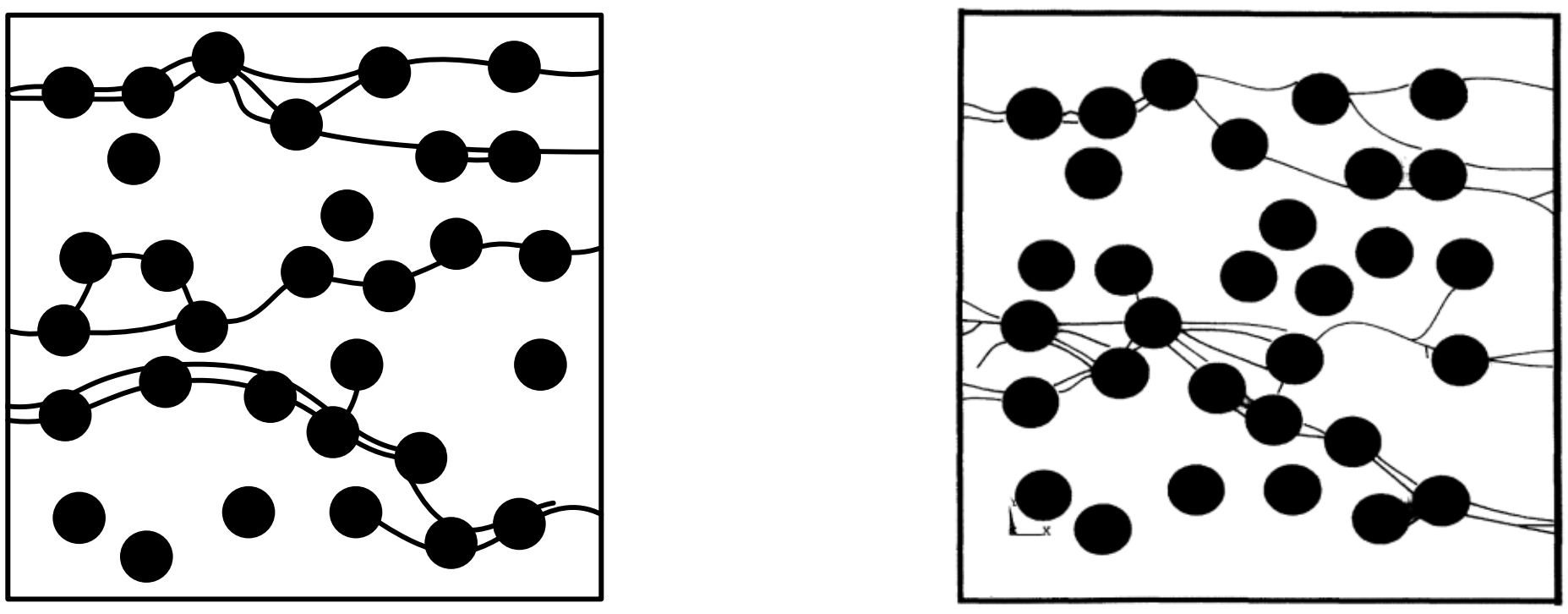

(a) (b)

Fig. 19 Summary of final crack paths under different load conditions: (a) simulation results of adaptive coupling of the PD-CCM and (b) experimental results [44].

# 6 Conclusion

In this study, an adaptive coupling of the PD-CCM driven by broken bond/strength criteria is presented to predict structural dynamic failure. Based on the proposed model, three benchmark examples are simulated: 1) dynamic crack branching of a pre-notched rectangular plate; 2) dynamic failure of a round disk with a center hole under the explosion load in the hole; and 3) dynamic fracture of a plate with random holes. Finally, the following conclusions are drawn from the simulation results:

(1) When the central difference method is used to solve the problem, as long as the expansion radius $r_p$ of the PD subdomain is greater than the radius $\delta$ of the PD horizon, the crack propagation will not be restricted by the PD subdomain.

(2) With the adaptive expansion of the PD subdomain, the CEs in the computational domain are converted into DEs for dynamic fracture simulation. During this process, the freedom of the entire system increases. The solution of the dynamic equation of the coupled model remains correct after reconfiguring the displacement vector and mass matrix at the nodes of the DEs.

(3) The adaptive coupling of the PD-CCM can effectively predict structural dynamic failure and minimize the area of the PD subdomains, reducing the calculation cost as much as possible.

# Acknowledgements

The authors gratefully acknowledge the financial support received from the National Natural Science Foundation (11872016) to complete this work.

# Reference

[1] Lubineau G, AzdoudY, Han F, Rey C, Askari A (2012) A Morphing strategy to couple non-local to local continuum mechanics. Journal of the Mechanics and Physics of Solids 60(6):1088-1102.

[2] Belytschko T, Black T (1999) Elastic crack growth in finite elements with minimal remeshing. International Journal for Numerical Methods in Engineering 45(5):601-620.

[3] Moës N, Dolbow J, Belytschko T (1999) A finite element method for crack growth without remeshing. International Journal for Numerical Methods in Engineering 46(1):131-150.

[4] Fan R, Fish J (2008) The rs-method for material failure simulations. International Journal for Numerical Methods in Engineering 73(11):1607-1623.

[5] Pandolfi A, Ortiz M (2012) An eigenerosion approach to brittle fracture. International Journal for Numerical Methods in Engineering 92(8):694-714.

[6] Francfort GA, Marigo JJ (1998) Revisiting brittle fracture as an energy minimization problem. Journal of the Mechanics and Physics of Solids 46(8):1319-1342.

[7] Bourdin B, Francfort GA, Marigo J-J (2000) Numerical experiments in revisited brittle fracture. Journal of the Mechanics and Physics of Solids 48(4):797-826.

[8] Ravi-Chandar K, Knauss WG (1984) An experimental investigation into dynamic fracture: III. On steady-state crack propagation and crack branching. International Journal of Fracture 26(2):141-154.

[9] Ravi-Chandar K (1998) Dynamic fracture of nominally brittle materials. International Journal of Fracture 90(1):83-102.

[10] Silling SA (2000) Reformulation of elasticity theory for discontinuities and long-range forces. Journal of the Mechanics and Physics of Solids 48(1):175-209.

[11] Silling SA, Lehoucq RB (2010) Peridynamic theory of solid mechanics. Advances in Applied Mechanics 44:73-168

[12] Silling SA, Epton M, Weckner O (2007) Peridynamic states and constitutive modeling. Journal of Elasticity 88(2):151-184.

[13] Kilic B, Agwai A, Madenci E (2009) Peridynamic theory for progressive damage prediction in center-cracked composite laminates. Composite Structures 90(2):141-151.

[14] Ha YD, Bobaru F (2010) Studies of dynamic crack propagation and crack branching with peridynamics. International Journal of Fracture 162(1):229-244.

[15] Le QV, Bobaru F (2018) Surface corrections for peridynamic models in elasticity and fracture, Computational Mechanics 61(4):499-518.

[16] Kilic B, Madenci E (2009) Coupling of peridynamic theory and finite element method. Journal of Mechanics of Materials and Structures 5(5):707-733.

[17] Agwai I. Guven E, Madenci E (2009) Damage prediction for electronic package drop test using finite element method and peridynamic theory. 59th Electronic Components and Technology Conference 565-569.

[18] Liu WY, Hong JW (2012) A coupling approach of discretized peridynamics with finite element method. Computer Methods in Applied Mechanics and Engineering 245:163-175.

[19] Li H, Zhang HW, Zheng Y, Ye H (2018) An implicit coupling finite element and peridynamic method for dynamic problems of solid mechanics with crack propagation. International Journal of Applied Mechanics 10:1850037.

[20] Shen F, Yu Y, Zhang Q, Gu XB (2020) Hybrid model of peridynamics and finite element method for static elastic deformation and brittle fracture analysis. Engineering Analysis With Boundary Elements 113:17-25.

[21] Ongaro G, Seleson P, Galvanetto U, Ni T (2021) Overall equilibrium in the coupling of peridynamics and classical continuum mechanics. Computer Methods in Applied Mechanics and Engineering.

[22] Galvanetto U, Mudric T, Shojaei A. M. Zaccariotto (2016) An effective way to couple FEM meshes and Peridynamics grids for the solution of static equilibrium problems. Mechanics Research Communications 76:41-47.

[23] Azdoud Y, Han F, Lubineau G (2013) A Morphing framework to couple non-local and local anisotropic continua. International Journal of Solids and Structures 50(9):1332-1341.

[24] Han F, Lubineau G, Azdoud Y, Askari A (2016) A Morphing approach to couple state-based peridynamics with classical continuum mechanics. Computer Methods in Applied Mechanics and Engineering 301:336-358.

[25] Han F, Liu SK, Lubineau G (2021) A dynamic hybrid local/nonlocal continuum model for wave propagation. Computational Mechanics 67(1):385-407.

[26] Azdoud Y, Han F, Lubineau G (2014) The Morphing method as a flexible tool for adaptive local/non-local simulation of static fracture. Computational Mechanics 54 (3):711-722.

[27] Wang YW, Han F, Lubineau G (2021) Strength-induced peridynamic modeling and simulation of fractures in brittle materials. Computer Methods in Applied Mechanics and Engineering 374:113558.

[28] Silling SA, Askari E (2005) A meshfree method based on the peridynamic model of solid mechanics. Computers & Structures 83(17/18):1526-1535.

[29] Wang YW, Han F, Lubineau G (2019) A hybrid local/nonlocal continuum mechanics modeling and simulation of fracture in brittle materials. Computer Modeling in Engineering and Sciences 121(2):399-423.

[30] Li ZB, Han F (2023) The peridynamics-based finite element method (PeriFEM) with adaptive continuous/discrete element implementation for fracture simulation. Engineering Analysis with Boundary Elements 146:56-65.

[31] Gao H (1993) Surface roughening and branching instabilities in dynamic fracture. Journal of the Mechanics and Physics of Solids 41(3):457-486.

[32] Silling SA, Demmie P, Warren TL (2007) Peridynamic simulation of high-rate material failure, 2007 ASME applied mechanics and materials conference.

[33] Han F, Li ZB (2022) A peridynamics-based finite element method (PeriFEM) for quasi-static fracture analysis. Acta Mechanica Solida Sinica 35(3):446-460.

[34] Song JH, Wang H, Belytschko T (2008) A comparative study on finite element methods for dynamic fracture. Computational Mechanics 42(2):239-250.

[35] Ha YD, Bobaru F (2011) Characteristics of dynamic brittle fracture captured with peridynamics. Engineering Fracture Mechanics 78(6): 1156-1168.

[36] Ha YD, Bobaru F (2010) Studies of dynamic crack propagation and crack branching with peridynamics. International Journal of Fracture 162(1):229-244.

[37] Bowden FP, Brunton JH, Field JE, Heyes AD (1967) Controlled fracture of brittle solids and interruption of electrical current. Nature 216:38-42.

[38] Döll W (1975) Investigations of the crack branching energy. International Journal of Fracture 11:184-186

[39] Banadaki MMD (2012) Numerical simulation of stress wave induced fractures in rock. International Journal of Impact Engineering 40:16-25.

[40] Trivino L, Mohanty B, Munjiza A (2009) Seismic radiation patterns from cylindrical explosive charges by analytical and combined finite-discrete element methods. In: Proceedings of the 9th International Symposium on Rock Fragmentation by Blasting, Fragblast 9:415-426.

[41] Kutter HK (1971) On the fracture process in blasting. International Journal of Rock Mechanics and Mining Sciences and Geomechanics Abstracts 8(3):181-202;

[42] Xie LX, Lu WB, Zhang QB, Jiang QH (2016) Damage evolution mechanisms of rock in deep tunnels induced by cut blasting. Tunnelling and Underground Space Technology 58:257-270.

[43] Al-Ostaz A, Jasiuk I (1997) Crack initiation and propagation in materials with randomly distributed holes. Engineering Fracture Mechanics 58(5-6):395-420.

[44] Ostoja-Starzewski M, Wang G (2006) Particle modeling of random crack patterns in epoxy plates. Probabilistic Engineering Mechanics 21(3):267-275.

[45] Ashby MF, Jones DRH (1980) Engineering materials 1: An introduction to their properties and applications. Oxford: Pergamon Press.